\documentclass[]{elsarticle} 
\usepackage{url}  
\usepackage{graphicx}  

\usepackage{booktabs} 
\usepackage{amsmath}
\usepackage{siunitx}
\usepackage{amsthm}
\usepackage{relsize}
\usepackage{amssymb}
\usepackage{algorithm}
\usepackage{booktabs}
\usepackage{verbatim} 
\usepackage{subfigure}
\usepackage{balance}
\usepackage{graphicx}
\usepackage{tikz}
\usepackage{algpseudocode}
\usepackage[bottom]{footmisc}
\usetikzlibrary{fit,positioning}
\usepackage[utf8]{inputenc}
\usepackage[english]{babel}
\usepackage{caption}
\usepackage{mathtools}
\usepackage{url}
\usepackage{hyperref}
\usepackage{multicol}
\usepackage{multirow}
\usepackage{xcolor}

\newtheorem{Dfn}{Definition}

\newcommand{\lanyu}[1]{}

\begin{document}
\begin{frontmatter}

\title{Towards Reliable Online Clickbait Video Detection: \\A Content-Agnostic Approach}
\author{Lanyu Shang}
\author{Daniel Zhang}
\author{Michael Wang}
\author{Shuyue Lai}
\author{Dong Wang}
\address{Department of Computer Science and Engineering\\
University of Notre Dame\\
Notre Dame, IN 46556}
\begin{abstract}
Online video sharing platforms (e.g., YouTube, Vimeo) have become an increasingly popular paradigm for people to consume video contents. Clickbait video, whose content clearly deviates from its title/thumbnail, has emerged as a critical problem on online video sharing platforms.  Current clickbait detection solutions that mainly focus on analyzing the text of the title, the image of the thumbnail, or the content of the video are shown to be suboptimal in detecting the online clickbait videos. In this paper, we develop a novel content-agnostic scheme, Online Video Clickbait Protector (OVCP), to effectively detect clickbait videos by exploring the comments from the audience who \lanyu{have} watched the video. Different from existing solutions, OVCP does not directly analyze the content of the video and its pre-click information (e.g., title and thumbnail). Therefore, it is robust against sophisticated content creators who often generate clickbait videos that can bypass the current clickbait detectors. We evaluate OVCP with a real-world dataset collected from YouTube. Experimental results demonstrate that OVCP is effective in identifying clickbait videos and significantly outperforms both state-of-the-art baseline models and human annotators. 
 
\end{abstract}

\begin{keyword}
Clickbait Video, Content Agnostic, Online Video Sharing, YouTube
\end{keyword}

\end{frontmatter}

\section{Introduction}
\label{sec:intro}

In the age of instant gratification, people are increasingly consuming more video contents from the Internet (e.g., YouTube\footnote{https://www.youtube.com}, Vimeo\footnote{https://vimeo.com/watch}) than cable networks~\cite{groupm2018}. The time people spend on online media is expected to surpass the time they spend on traditional TV worldwide in 2019 \cite{recode2019}. For example, YouTube has over a billion users covering almost one-third of the Internet population and reaches billions of views per day \footnote{https://www.youtube.com/yt/about/press/}. 
An online video usually includes \lanyu{the} title, \lanyu{the} thumbnail, and the video content. The title and thumbnail are visible to the viewers ~\emph{before} they click the video and are found to be the crucial factors that may attract users to click and watch a video. 
\lanyu{Clickbait} video, whose content clearly deviates from its title/thumbnail, is generated to entice viewers to click \lanyu{the} video and boost the viewership of the video consequently \cite{bartl2018youtube}.
However, the spread of clickbait videos not only wastes the time of viewers but also decreases the trustworthiness of video-sharing platforms.
In this paper, we focus on the problem of reliably detecting clickbait videos \lanyu{on} online video-sharing platforms.

\begin{figure}[!htb]
    \centering
    \subfigure[][Clickbait]{
        \centering
        \includegraphics[width=0.4\linewidth]{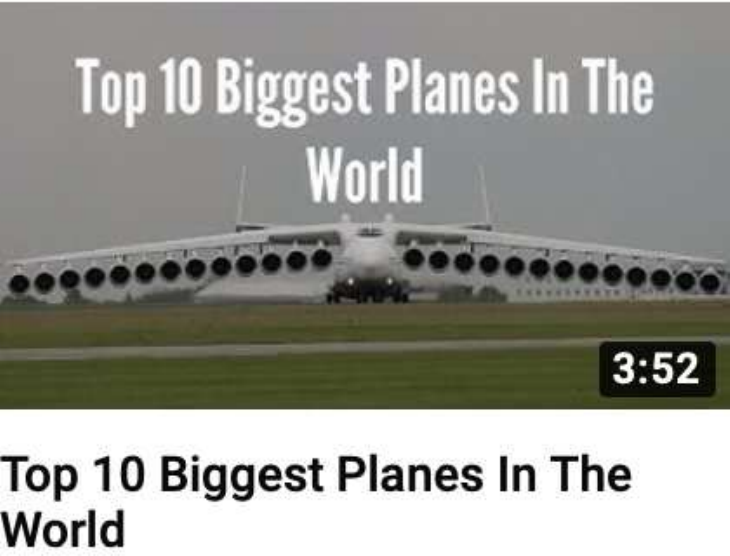}
        \label{fig:example_title_clickbait}
    }
    \hspace{0.1in}
    \subfigure[][Non-clickbait]{
        \centering
        \includegraphics[width=0.4\linewidth]{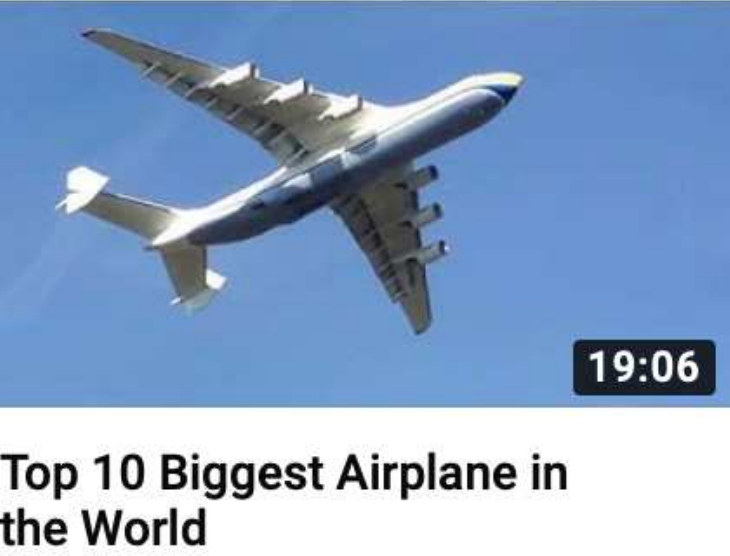}
        \label{fig:example_title_nonclickbait}
     }
    \caption{Examples of Clickbait and Non-clickbait Video with Similar Titles}
    \label{fig:example_title}
\end{figure}

The detection of clickbait videos requires a careful investigation of the relationship between the title/thumbnail and the video content. This problem cannot be fully addressed by current content-based solutions which only focus on the text of the title \cite{rony2017diving, wang2015social}, the image of the thumbnail \cite{zeiler2014visualizing, zhang2018opinion}, or the content of the video \cite{papadopoulou2017web, zhang2018streamguard}. 
For example, several text-based clickbait detection techniques have been developed to identify clickbait from social media posts (e.g., the clickbait news headline detection using word embeddings \cite{rony2017diving}, \lanyu{the} clickbait tweets detection using the linguistic feature analysis \cite{potthast2016clickbait}).
However, those solutions cannot be adopted to address the video clickbait detection problem because the content of the title may not be a reliable indicator to identify a clickbait video. For example, both videos shown in Figure \ref{fig:example_title} share the same title. The video shown in Figure \ref{fig:example_title_clickbait} is a clickbait because it does not deliver the content the viewers expect to see (e.g., a plane with 13 engines on each wing as shown in its thumbnail). In contrast, Figure \ref{fig:example_title_nonclickbait} is a non-clickbait video and the video does present the ten largest airplanes in the world including the one shown in the thumbnail.

\begin{figure}[!htb]
    \centering
    \subfigure[][Clickbait]{
        \centering
        \includegraphics[width=0.4\linewidth]{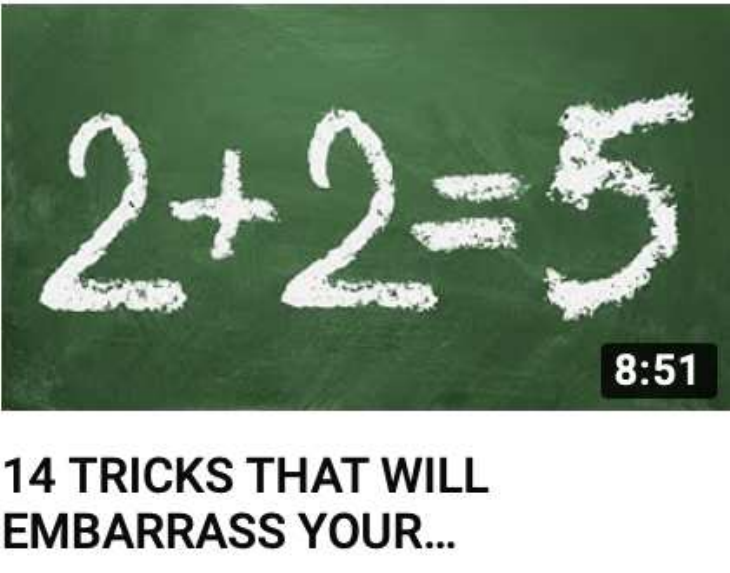}
        \label{fig:example_thumbnail_clickbait}
    }
    \hspace{0.1in}
    \subfigure[][Non-clickbait]{
        \centering
        \includegraphics[width=0.4\linewidth]{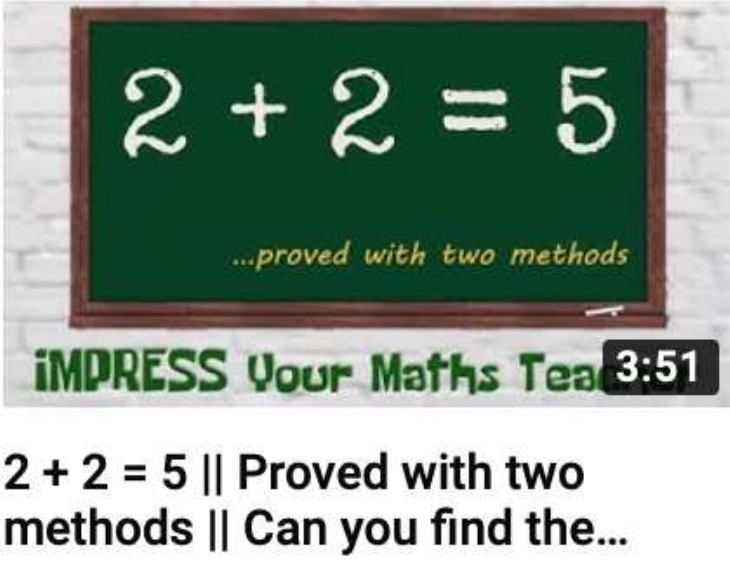}
        \label{fig:example_thumbnail_nonclickbait}
    }
    \caption{Examples of Clickbait and Non-clickbait Video with Similar Thumbnails}
    \label{fig:example_thumbnail}
\end{figure}

Similarly, image-based approaches that only focus on thumbnail features (e.g., convolutional method \cite{zeiler2014visualizing}, self-consistency \cite{huh2018fighting}) cannot solve the video clickbait detection problem. 
For example, Figure \ref{fig:example_thumbnail} shows two videos with very similar thumbnails. The video \ref{fig:example_thumbnail_clickbait} in Figure \ref{fig:example_thumbnail} is a clickbait because the video only presents some irrelevant tricks, such as how to plot letters ``L'', ``O'', ``V'', ``E'' with math functions, and never explains the equation $2+2=5$. 
In contrast, the video shown in Figure \ref{fig:example_thumbnail_nonclickbait} is a non-clickbait as its content explicitly discusses some tricky methods to derive the equation $2+2=5$ (e.g., $2\times 0+2 \times 0=5\times 0$).

Moreover, video authentication and verification solutions have been developed to identify fake online videos \cite{papadopoulou2017web} and AI-generated videos \cite{li2018ictu}. However, these solutions only focus on the video content itself and ignore the information exposed to viewers before the video is clicked (e.g., title \lanyu{and} thumbnail). For example, an old movie can be uploaded under a title/thumbnail \lanyu{that is} identical or similar to a recently released one. Such a clickbait video can easily bypass the video-based detection systems as the video content is real \cite{fagan2018}. Therefore, a reliable online clickbait video detection tool that explicitly considers the relationship between the title/thumbnail and the video content has yet to be developed.

In this paper, we develop a novel content-agnostic scheme, \textit{Online Video Clickbait Protector (OVCP)}, to effectively detect clickbait videos whose titles/thumbnails are deviating from the video content. OVCP is content-agnostic in the sense that it does not analyze the content information of a video (i.e., title, thumbnail, video clips). Instead, OVCP investigates the comments and discussions from online users who \lanyu{have} watched the video. Our approach is motivated by the observation that users usually make complaints or sarcastic comments when they watch clickbait videos (e.g., ``My body is crying for help watching this video'' in a clickbait video titled ``28 SIGNS YOUR BODY IS CRYING FOR HELP''). 
Additionally, users are observed to be less interactive in the discussion of clickbait videos than non-clickbait ones given the fact that the content of clickbait videos often appears to be tedious to the users~\cite{marketing2018}.  The OVCP scheme explores the topological and semantic structure of the user comment networks and the latent linguistic features of user comments to identify the unique characteristics of clickbait videos. OVCP is then integrated with a set of state-of-the-art supervised classifiers to detect clickbait videos. 

To the best of our knowledge, OVCP is the first content-agnostic approach to address the online clickbait video detection problem. 
OVCP is robust against \lanyu{sophisticated content creators} who often generate clickbait videos that can bypass the current clickbait detectors. The evaluation results on a real-world dataset collected from YouTube show that our OVCP scheme accurately detects clickbait videos and significantly outperforms state-of-the-art baselines and human annotators.

\section{Related Work}

\subsection{Clickbait Detection}

Our work bears some resemblance with literature on clickbait detection on social media platforms. The term clickbait was coined as ``exaggerated headlines whose main motive is to mislead the reader to click on them" \cite{agrawal2016clickbait}. The problem of clickbait has been widely studied and many solutions were provided~\cite{potthast2016clickbait,potthast2018crowdsourcing,anand2017we}. The first automatic clickbait detector was proposed in \cite{potthast2016clickbait}, where a set of handcrafted features (e.g., bag-of-words, n-grams and number of hashtags) have been selected to train a clickbait classifier.  The follow-up solutions improve this feature engineering approach by developing deep neural network based approaches to automatically detect clickbaits without cherry picking the features \cite{anand2017we,thomas2017clickbait}. Unfortunately, these schemes rely on the analysis of the textual content (i.e., titles and descriptors) to detect clickbaits and cannot effectively address our problem because both clickbait videos and non-clickbait videos can share very similar titles and descriptors as shown in Figure \ref{fig:example_title}.


We found there exist very few clickbait detection solutions that focus on video-based clickbaits. In a very recent work,   Qu \emph{et al.} developed a crowdsourcing-based approach where human annotators are invited to label whether the thumbnail of a video is clickbait or not \cite{qutowards}. This work requires a significant amount of human labor and can be time-consuming if the labeled dataset is large. The most relevant work was proposed by Zannettou \textit{et al.}, who developed a deep neural network model  to combine the features of the thumbnails and the statistical features of users' comments \cite{zannettou2018good}. However, this approach relies on the direct analysis of the thumbnails which has been shown to be an unreliable indicator of clickbait videos (Figure \ref{fig:example_thumbnail}).
Moreover, the scheme assumes all the videos posted by the same user were clickbaits which is an oversimplified assumption for the real-world scenarios where sophisticated users exist. In contrast, the proposed OVCP scheme is the first \emph{content-agnostic} approach to address the clickbait video detection problem on video sharing platforms.

\subsection{Misinformation Detection}
The spread of misinformation on online social media has received a significant amount of attention in recent years \cite{zhang2018towards,vo2018rise,R33,wang2014using,wang2013recursive,zhang2018scalable}.  There exist two categories of misinformation detection problems that are related to ours. The first category is called \emph{fake news detection} where \emph{textual content} (e.g., news articles and social media posts) is analyzed to verify whether the news statement is truthful or not.  For example, Vo \textit{et al.} developed a fake news detection scheme that can identify a group of users who actively debunk fake information on social media and recommend fact-checking URLs posted from these users \cite{vo2018rise}. Wang \emph{et al.} developed an estimation theoretical scheme that identifies truthful online social media posts by explicitly considering the reliability of data sources and source dependency~\cite{R1}. The second category is commonly referred to as ``image forgery detection" schemes. For example, Zhang \emph{et al.} developed a novel fauxtography detector that can effectively track down misleading images on social media (e.g., Twitter, Reddit)  \cite{zhang2018fauxbuster}. Huynh-Kha \emph{et al.} developed an image forgery detection scheme that can detect whether an image is manually edited by copy-move, splicing or both in the same image \cite{huynh2016robust}.

However, the above solutions are insufficient to address the clickbait video detection problem in this paper because sophisticated uploaders can often create clickbait videos by using credible titles and non-edited images to bypass the detection systems~\cite{zannettou2018good}. In this work, we propose a more robust content-agnostic approach that leverages the audience's comments to detect clickbait videos rather than \lanyu{relies} on the analysis of textual descriptors or thumbnails.


\subsection{Network Embedding for Comment Feature Extraction}

Extracting topological and semantic information from the comment network is a key technique used in the proposed OVCP scheme. Various relevant network embedding techniques have been proposed \cite{wang2018shine,grover2016node2vec,perozzi2014deepwalk,huang2017label,zhang2018risksens}. For example, DeepWalk uses short random walks to learn the latent representations to describe the topological structure of a network  \cite{perozzi2014deepwalk}.  Node2vec designs a biased random walk procedure to learn the representation of a network that maximizes the likelihood of preserving network neighborhoods of a node \cite{grover2016node2vec}. These classical methods focus on capturing the topological structure of the nodes while ignoring the attributes of the edge. In a more recent study, Wang \emph{et al.} developed SHINE, a network embedding technique that can capture the positive/negative emotions of the edge in a network \cite{wang2018shine}. Huang \emph{et al.} further extended the effort by proposing LANE, a general network embedding technique designed for attributed networks \cite{huang2017label}. Compared to SHINE, LANE is able to capture general categorical attributes on the edges of a network. Different from the above methods, our work encodes the unique sentiment and endorsement features of the user comment network for the purpose of detecting clickbait videos.


\section{Problem Definition}
\label{sec:problem}
In this section, we present the online clickbait video detection problem. First, we define a few key terms that will be used in the problem formulation. 

\begin{figure*}[!htb]
 \centering
    \subfigure[][Title and Thumbnail]{
         \centering
         \includegraphics[width=0.35\textwidth]{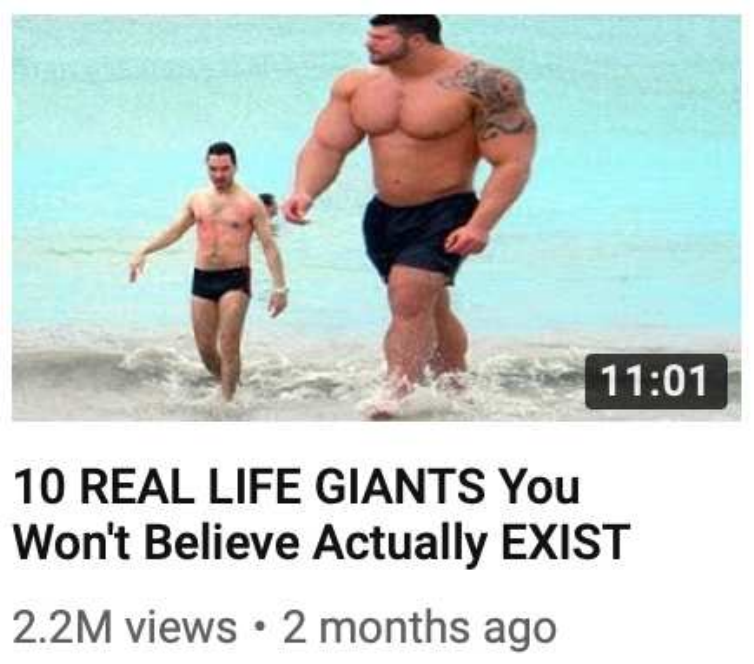}
         \label{fig:example-title-thumbnail}
     }
     \hspace{0.1in}
     \subfigure[][Description]{
         \centering
         \includegraphics[width=0.57\textwidth]{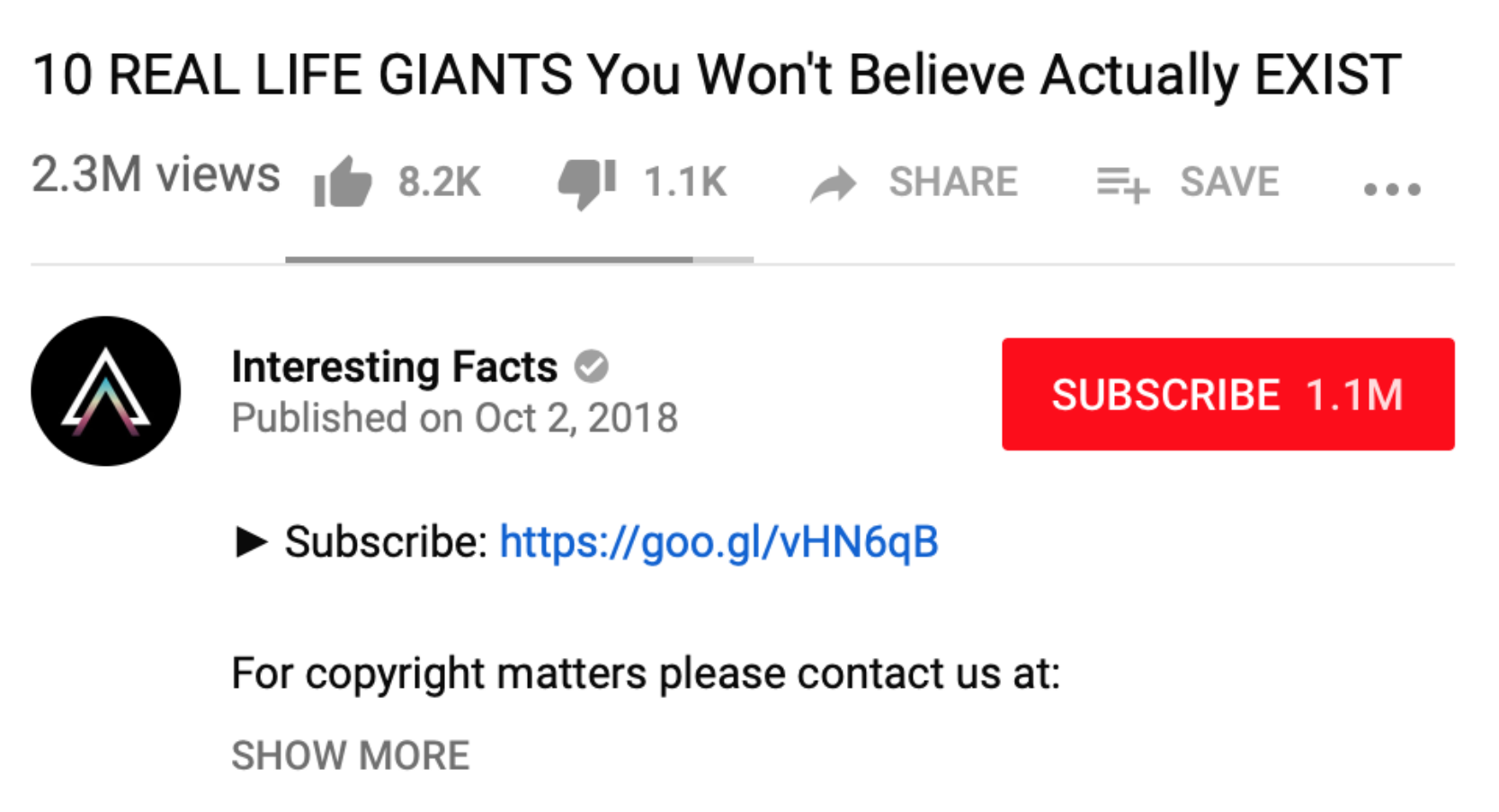}
        \label{fig:example-description}
     }
     \subfigure[][Sample Comments]{
         \centering
         \includegraphics[width=0.8\textwidth]{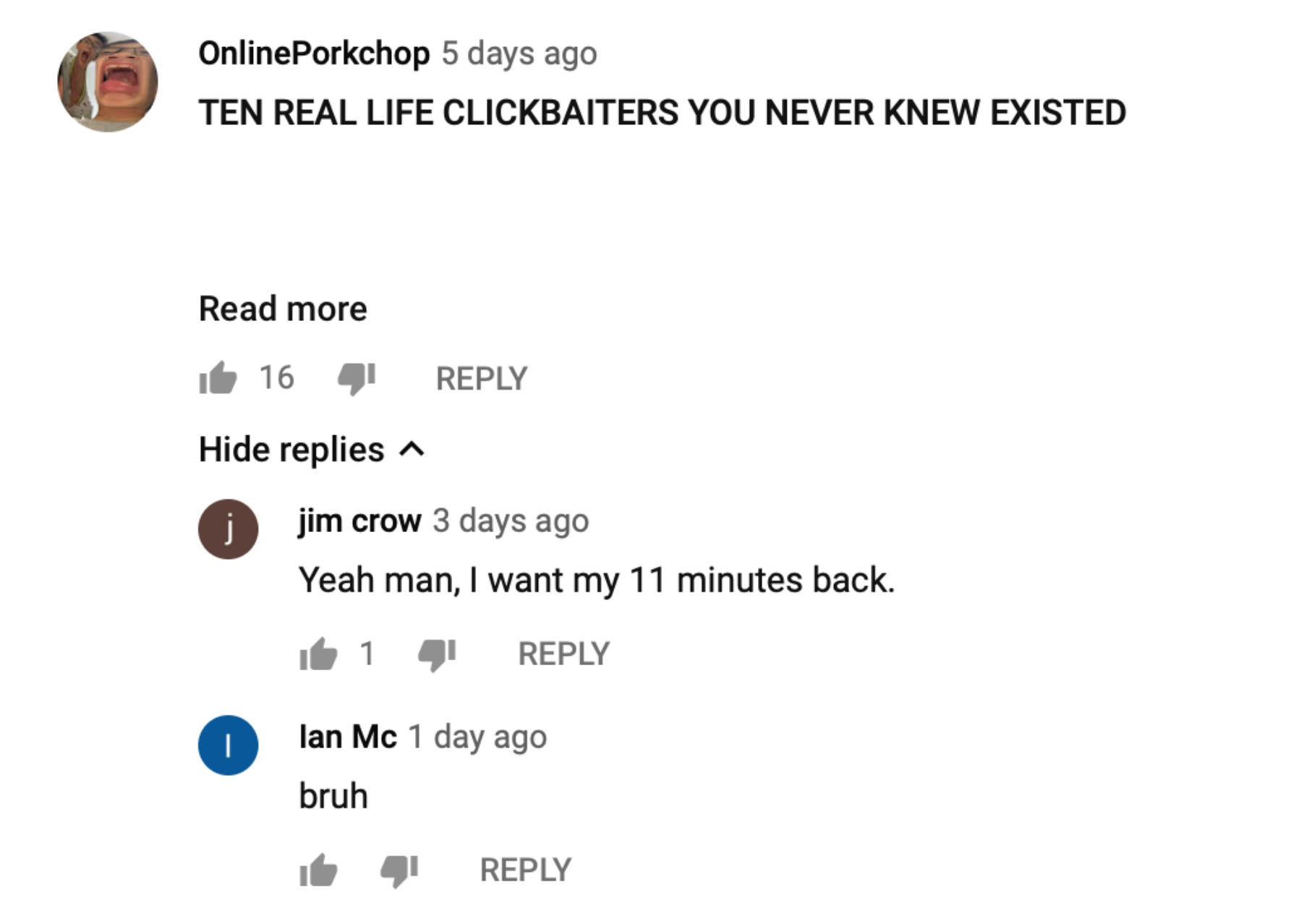}
        \label{fig:example-comments}
     }
     \caption{Example of an Online Video and Its Components}
     \label{fig:example1}
 \end{figure*}

\begin{Dfn}
\textbf{Video ($V_i$):} it is a publicly available video instance uploaded to the online sharing platforms (Figure~\ref{fig:example1}). Each video, denoted as $V_i$, contains five elements: title ($T_i$), thumbnail ($M_i$), description ($D_i$), comments ($C_i$), and the ground truth label ($z_i$) of the video. Formally, $V_i$ is denoted as a quintuple $V_i = (T_i, M_i, D_i, C_i, z_i)$. Additionally, we define a video set of $N$ videos as $\mathcal{V} = \{ V_1, V_2, \cdots, V_N\}$. 
\end{Dfn}

\begin{Dfn}
\textbf{Title ($T_i$)}: it is a piece of brief textual information provided by the video uploader to describe the video content. An example of a video title is shown in Figure \ref{fig:example-title-thumbnail}.
\end{Dfn}

\begin{Dfn}
\textbf{Thumbnail ($M_i$):} it is an image that provides a visual description of the video content. YouTube video creators are allowed to choose a thumbnail generated automatically by the YouTube algorithm, which is typically a frame from the video, or to upload a customized image. An example of a video thumbnail is shown in Figure \ref{fig:example-title-thumbnail}.
\end{Dfn}

\begin{Dfn}
 
\textbf{Description ($D_i$):} it includes the description provided by the video creator, and the video's meta information (e.g., number of views, number of likes). An example of a video description is shown in Figure \ref{fig:example-description}. 
\end{Dfn}

\begin{Dfn}
\textbf{Comments ($C_i$):} they include all the comments and reviews of a video from its viewers. An example of the comment section of a video is shown in Figure \ref{fig:example-comments}.
\end{Dfn}


\begin{Dfn}
\textbf{Clickbait Video (labeled as ``true"):} a video is defined as \textit{clickbait} if its title and/or thumbnail is intentionally crafted to attract viewers' attention, and entice them to click the accompanied video whose content clearly deviates from the title and/or thumbnail.
\end{Dfn}

\begin{Dfn}
 
\textbf{Non-clickbait Video (labeled as ``false"):} any video that is not in the clickbait video category.
 
\end{Dfn}

Based on the above definitions, the clickbait video detection is a \emph{binary classification problem} that targets at classifying each video into two categories, i.e., clickbait or non-clickbait. The problem is formulated as:
\begin{equation}
    \arg \max_{\Tilde{z_i}} Pr(\Tilde{z_i} = z_i|V_i) ,~  \forall 1 \le i \le N
\end{equation}
\noindent
where $\Tilde{z_n}$ denotes the estimated label for video $V_n$.

\section{Solution}
\label{sec:solution}
\begin{figure}[!htb]
    \centering
    \includegraphics[width=\linewidth]{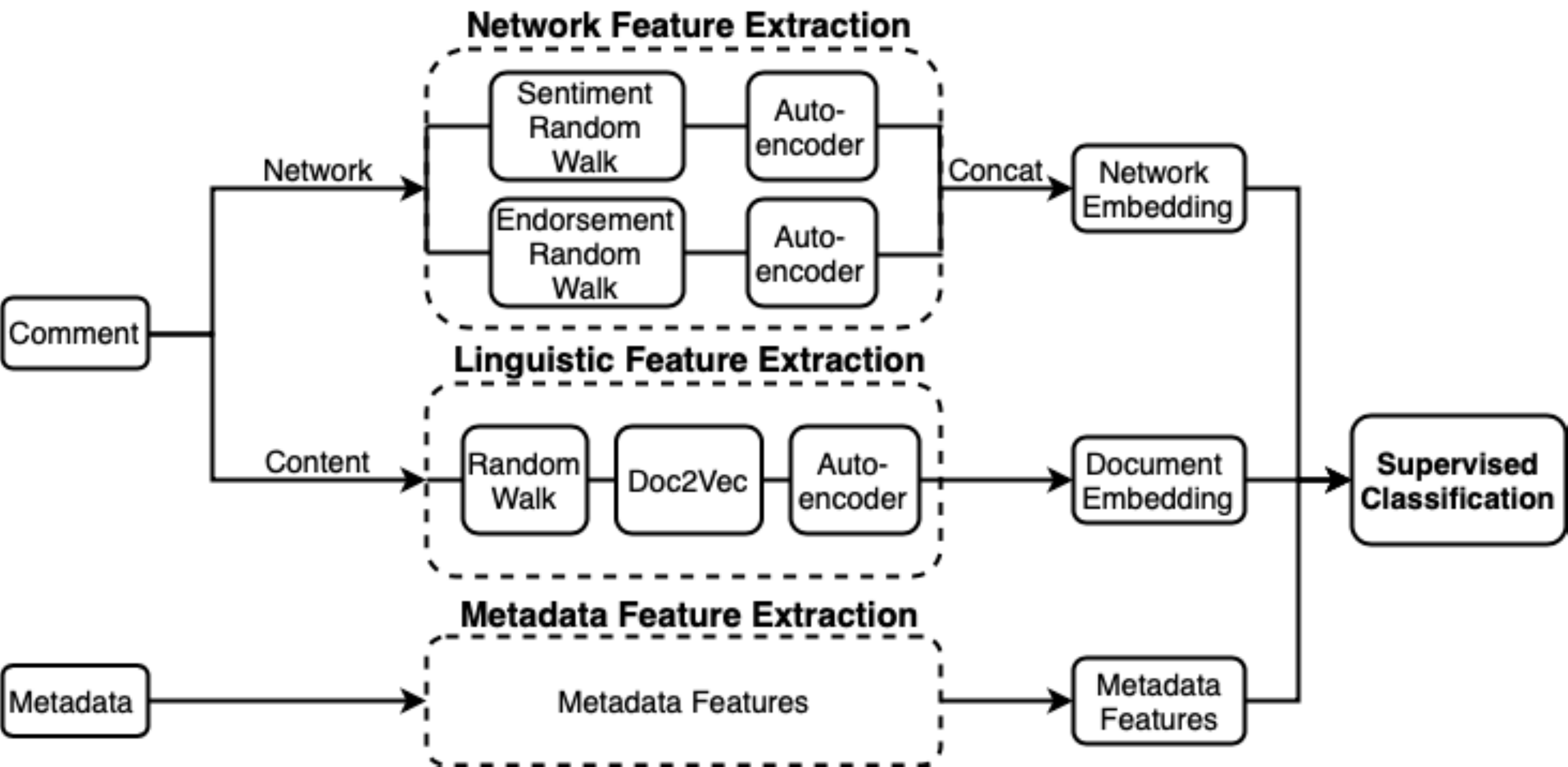}
    \caption{An Overview of OVCP}
    \label{fig:architecture}
\end{figure}

In this section, we present the \textit{Online Video Clickbait Protector (OVCP)} to address the problem formulated in the previous section. Our scheme consists of four major components as shown in Figure \ref{fig:architecture}. This first component is a \textit{Network Feature Extraction} module that extracts topological and semantic features from the audience interactive discussions and comments of a video.  The second component is a \textit{Linguistic Feature Extraction} module that learns the vector representations of user comments and extracts linguistic features of the comments (e.g., document embeddings). The third component is a \textit{Metadata Feature Extraction} module that extracts auxiliary metadata features of the video (e.g., number of views, number of likes).  The last component is a \textit{Supervised Classification} module that identifies the clickbait videos using the features extracted from the first three components.  

\subsection{Network Feature Extraction} \label{sec:network}

The network feature extraction component in OVCP is designed to capture characteristics from the audience's comments of an online video. We observe that clickbait and non-clickbait videos are different in terms of topological features (e.g., structure of comment threads) and semantic features (e.g., sentiments, endorsements) of audience's comments. To effectively capture both topological and semantic features, the network feature extraction component constructs a comment network, records the network attributes by a Random Walk scheme, and learns the vector representation via an autoencoding approach.

\subsubsection{Comment Network Construction for Individual Video}

We first construct a network of user comments that represents the topological structure and semantic attributes/features of the user comments. A \emph{comment network $\mathbf{G}$} for each video is defined as a directed graph $\mathbf{G} = (\mathbf{V}, \mathbf{E})$, where $\mathbf{V}$ is the set of nodes and $\mathbf{E}$ is the set of directed edges between nodes. We define a source node $s \in \mathbf{V}$ to be the description of the video, and other nodes (i.e., $v \in \mathbf{V}, v \neq s$) to be all comments of the video. Each edge $(v,v') \in \mathbf{E}$ denotes a reply from comment $v$ to comment $v'$.

Different from other trending social media platforms (e.g., Reddit) that have a multi-level comment structure, YouTube only allows a two-level commentary mechanism, namely a top-level comment and its replies (i.e., second-level comments to the top-level comment). In other words, each comment thread is recorded as the top-level comment followed by all of its replies in a flat structure.  To retrieve the comment network structure, in each comment thread, we connect the top-level comment node to the source node $s$ and direct all second-level comment nodes to their top-level comment node. When a specific user (e.g., a username following a ``+'' or ``@'' ) is mentioned in a second-level comment, the comment node is redirected to the latest comment node of the mentioned user in the same thread. Note that a comment can receive any number of replies but can only reply to one comment. In other words, each node $v$ can have any number of incoming edges but only one outgoing edge.

\begin{figure*}[!htb]
\centering
     \subfigure[][Clickbait]{
         \centering
         \includegraphics[width=0.42\textwidth]{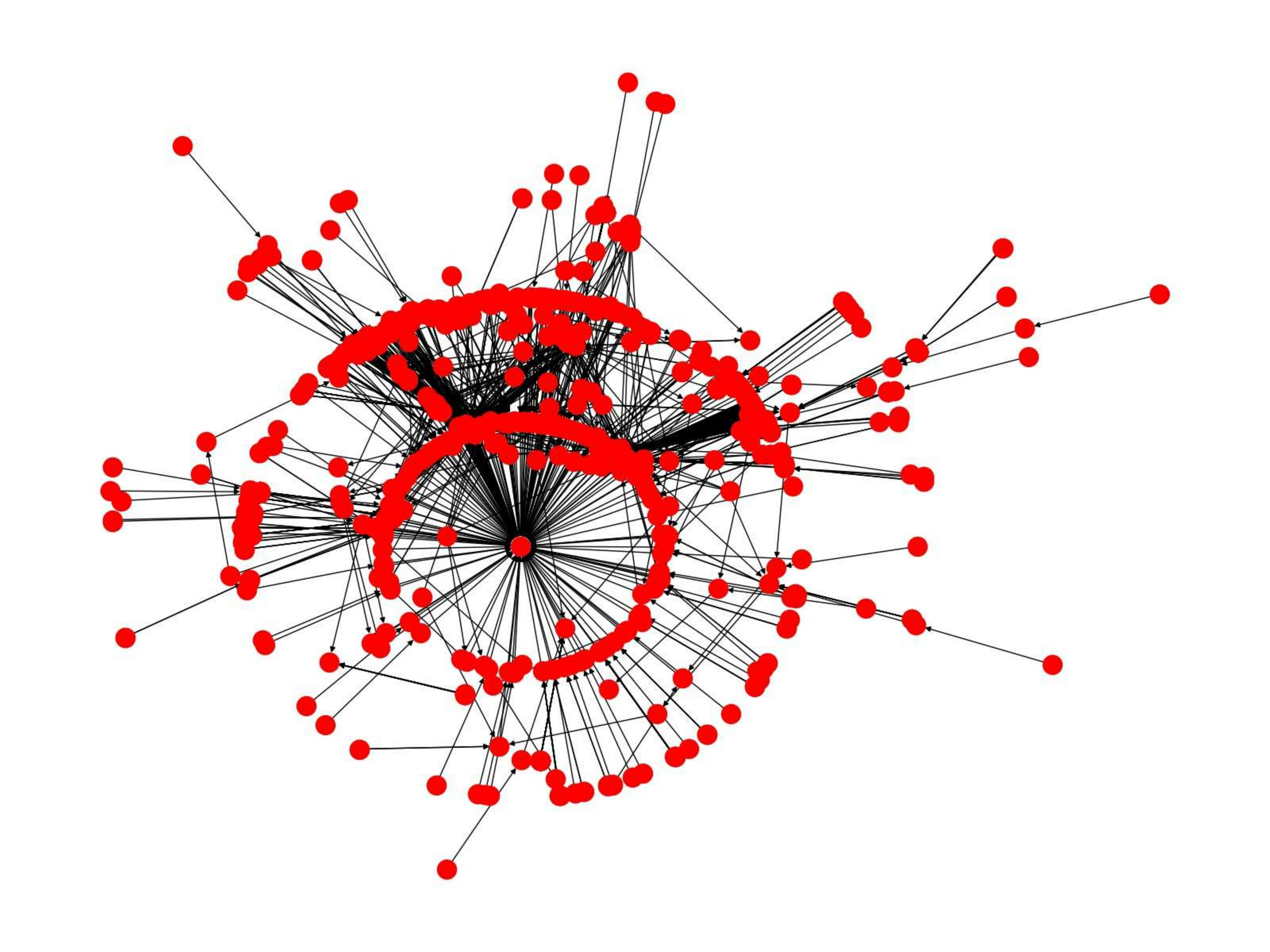}
         \label{fig:network-clickbait}
     }
     \subfigure[][Non-clickbait]{
         \centering
         \includegraphics[width=0.42\textwidth]{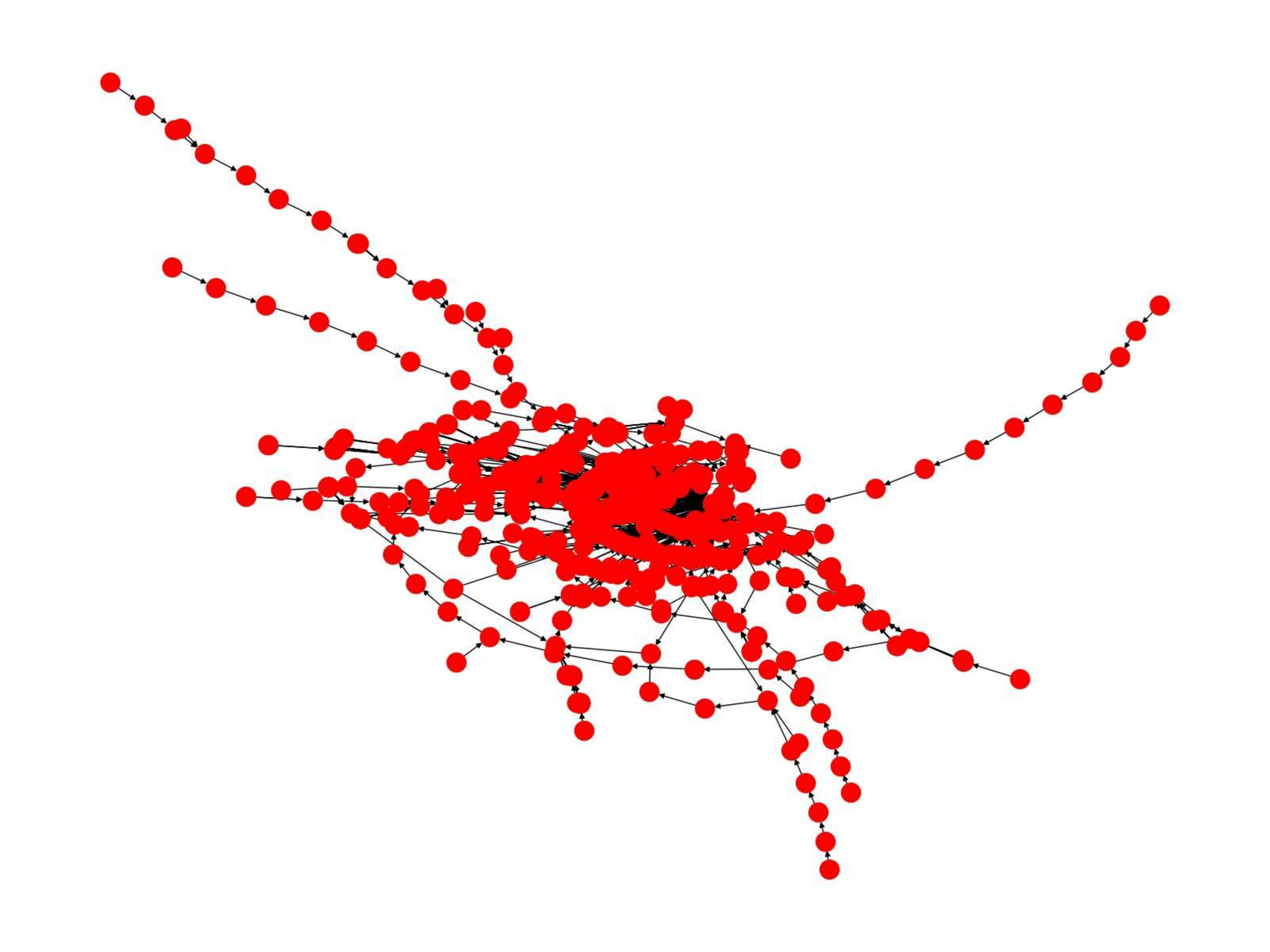}
        \label{fig:network-nonclickbait}
     }
     \caption{Examples of the Comment Network Structure  for Clickbait and Non-clickbait Videos  Note: for better visualization, we only keep threads with more than one comments in the plot.}
     \label{fig:network}`
 \end{figure*}

Figure \ref{fig:network} demonstrates the networks constructed from the comments of a clickbait video and a non-clickbait video. We observe that the comment network of a clickbait video has a few ``hub" comment nodes that receive more replies than normal comment nodes. For example, a comment like ``Who else come for the thumbnail?'' receives a large number of replies from other viewers who suffer from the same clickbait video. We also observe that non-clickbait videos contain longer comment threads since users are more likely to discuss the topic presented in the video through interactive replies if the video is not a clickbait.

\subsubsection{Semantic Feature Extraction with Random Walks}

After the comment network $\mathbf{G}$ of a video is constructed, we further extract the semantic features (i.e., sentiment and endorsement) from the comment network. We observe that such semantic features of a comment can sometimes reveal the user's attitude and behavior towards a possible clickbait video. Therefore, it makes sense to incorporate them into our OVCP scheme.  In particular, we adopt a Random Walk (RW) algorithm \cite{grover2016node2vec} in the constructed network to capture these two semantic features.

A random walk $RW(M, K)$ scheme often traverses a graph $M$ times while limiting the number of steps in each traverse path to be at most $K$ \cite{perozzi2014deepwalk}. In the OVCP scheme, we trace not only the reply direction of a comment node but also the attributes of the comment node. Intuitively, in each traversal of the comment network $\mathbf{G}$, we randomly select a comment node and record its path until it reaches the source node or the length of the path reaches $K$. 
We define the two attitude paths below to track the semantic features (i.e., sentiment and endorsement) of each comment node in the RW process.  In particular, we define two semantic features as attributes of each comment node $v$: i) \emph{sentiment attribute $\alpha_{s}(v)$} is defined as the \emph{polarity} score extracted by a sentiment analysis tool TextBlob\footnote{https://textblob.readthedocs.io/}, and ii) \emph{endorsement attribute $\alpha_{e}(v)$} is defined as the number of likes a comment received from other users.

\begin{Dfn}
\textbf{Sentiment Path ($RW_{s}$):} is the random walk process that traverses the graph $\mathbf{G}$ from a randomly selected comment node $v_0$ and records the \textit{sentiment} attribute $\alpha_{s}$ of each comment node on the path. In each step, the random walk process follows the direction to the next comment node that the current comment node replies to. Formally, for the $m^{th}$ walk in the process, $RW_{s}(m) = \{RW_{s}(m, 0), RW_{s}(m, 1), \cdots, RW_{s}(m, K-1)\}$ where $RW_{s}(m, k) = \alpha_{s}(v_k)$ represents the sentiment attribute of the $k^{th}$ node $v_k$ on the path. 
\end{Dfn}

\begin{figure*}[!htb]
\centering
     \subfigure[][Clickbait]{
         \centering
         \includegraphics[width=0.42\textwidth]{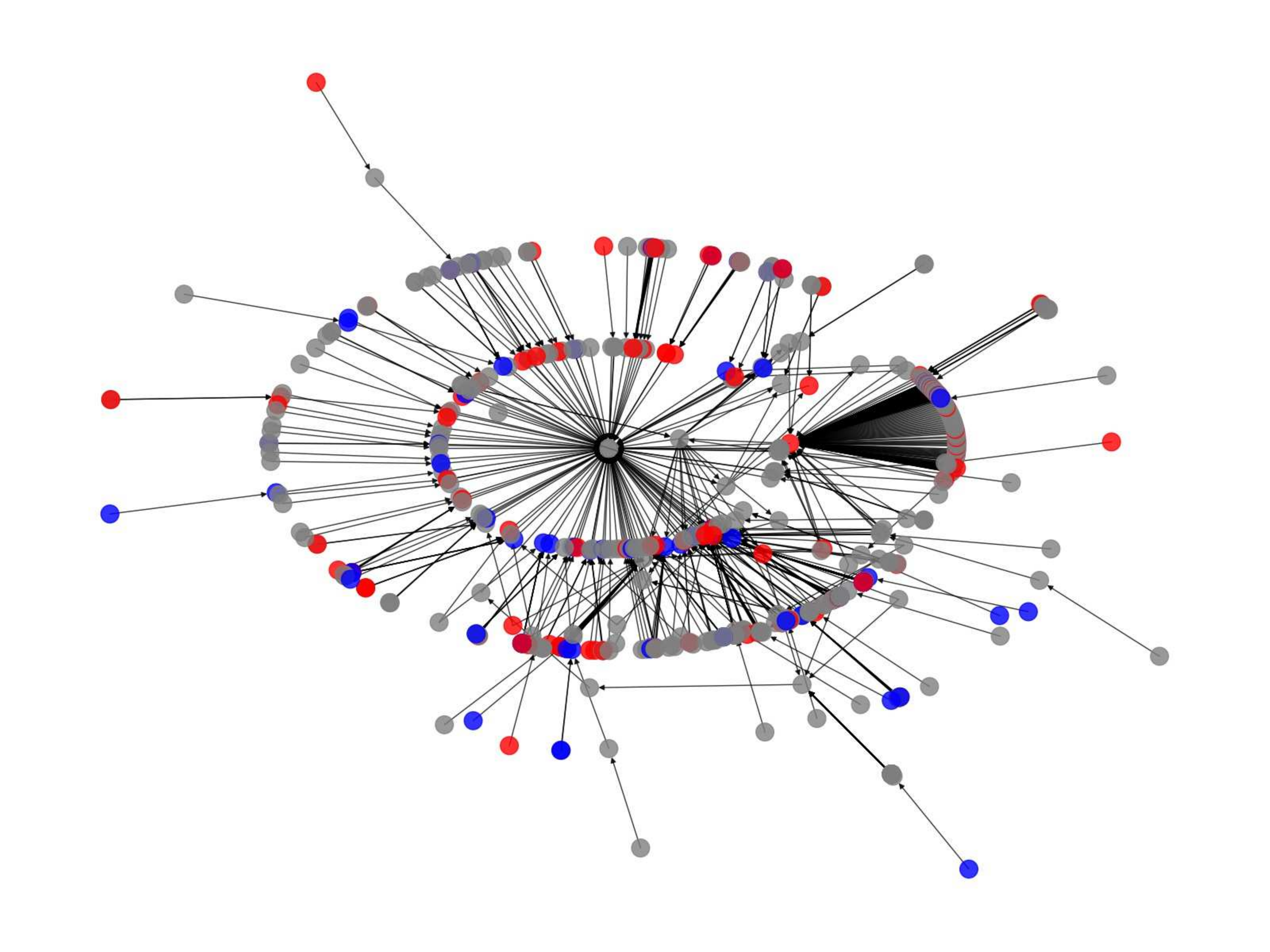}
         \label{fig:network-clickbait-color}
     }
     \subfigure[][Non-clickbait]{
         \centering
         \includegraphics[width=0.42\textwidth]{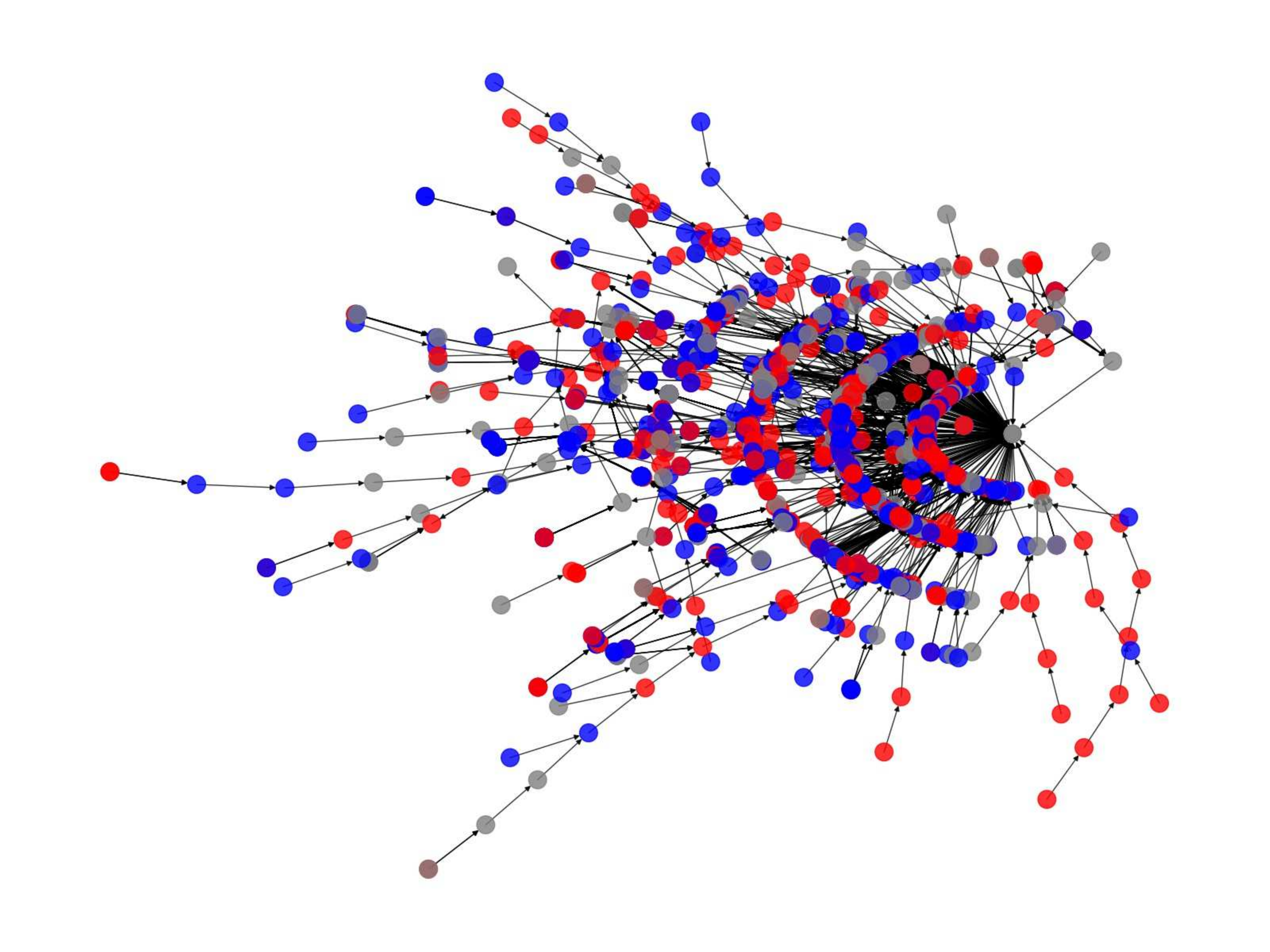}
        \label{fig:network-nonclickbait-color}
     }
     \caption{Sentiment Feature Path. The color of each node represents the sentiment attribute of the corresponding comment, i.e., \textit{red} - positive, \textit{blue} - negative, \textit{grey} - neutral.}
     \label{fig:network-color}
 \end{figure*}

The Sentiment Path $RW_{s}$ captures the sentiment feature of the comment network. Figure \ref{fig:network-color} demonstrates an example of the sentiment attribute in the comment network.  We observe that the comment sentiments from non-clickbait videos are more diverse than clickbait videos. This is because the comments from viewers of non-clickbait videos primarily focus on the content of the video and usually reflect the diversified attitudes (e.g., like vs. dislike) towards the video. Another reason is that viewers often leave a clickbait video immediately without making any comment. Such rational user behavior reduces not only the number of comments a clickbait video receives but also its likelihood of getting diversified comments.   

\begin{Dfn}
\textbf{Endorsement Path ($RW_{e}$):} is the random walk process that traverses the graph $\mathbf{G}$ from a randomly selected comment node $v_0$ and records the \textit{endorsement} attribute $\alpha_{e}$ of each comment node on the path. In each step, the random walk process follows the direction to the next comment node that the current comment node replies to. Formally, for the $m^{th}$ walk in the process, $RW_{e}(m) = \{RW_{e}(m, 0), RW_{e}(m, 1), \cdots, RW_{e}(m, K-1)\}$ where $RW_{e}(m, k) = \alpha_{e}(v_k)$ represents the endorsement attribute of the $k^{th}$ node $v_k$ on the path. 
\end{Dfn}

\begin{figure*}[!htb]
\centering
     \subfigure[][Clickbait]{
         \centering
         \includegraphics[width=0.42\textwidth]{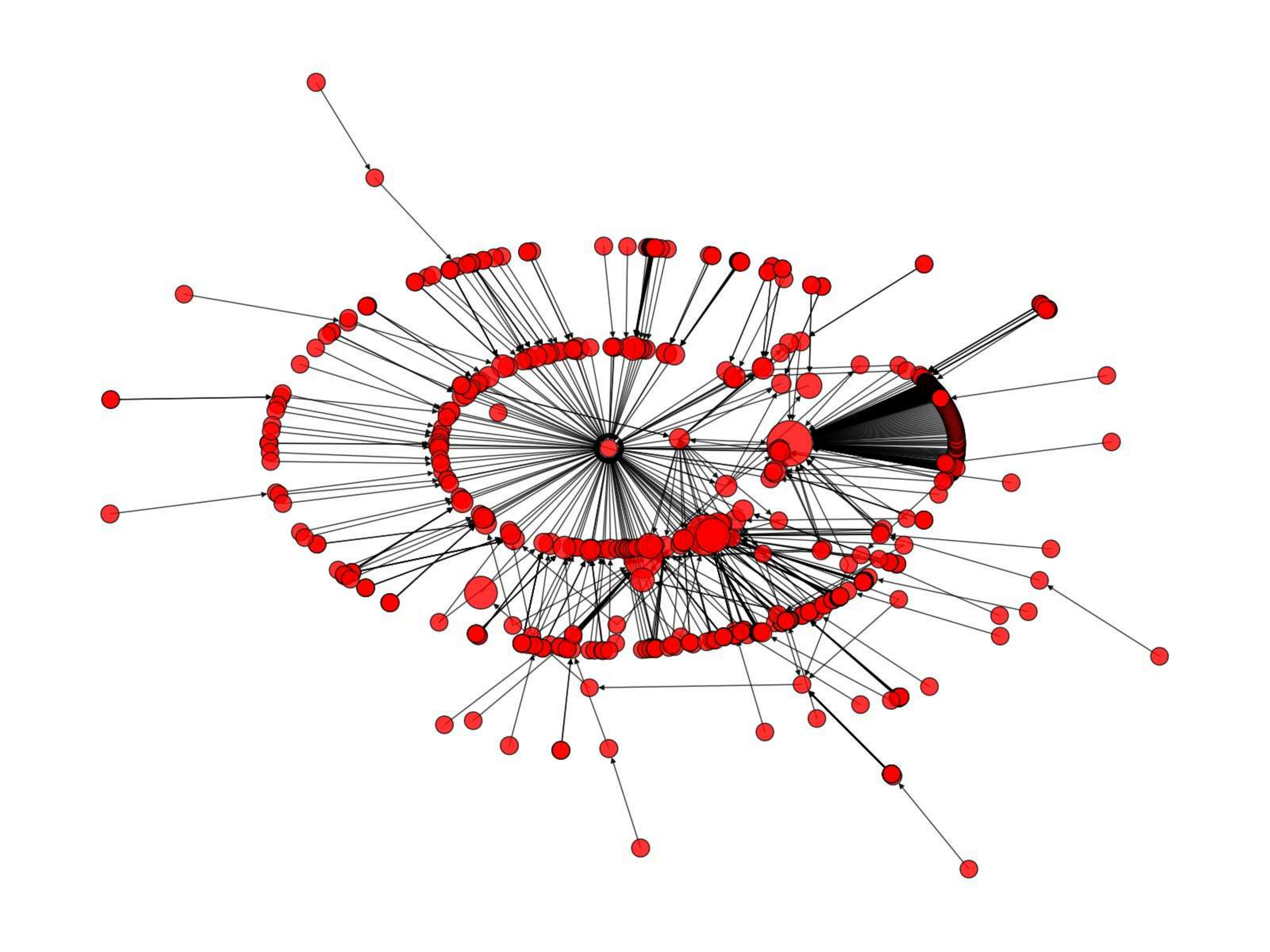}
         \label{fig:network-clickbait-size}
     }
     \subfigure[][Non-clickbait]{
         \centering
         \includegraphics[width=0.42\textwidth]{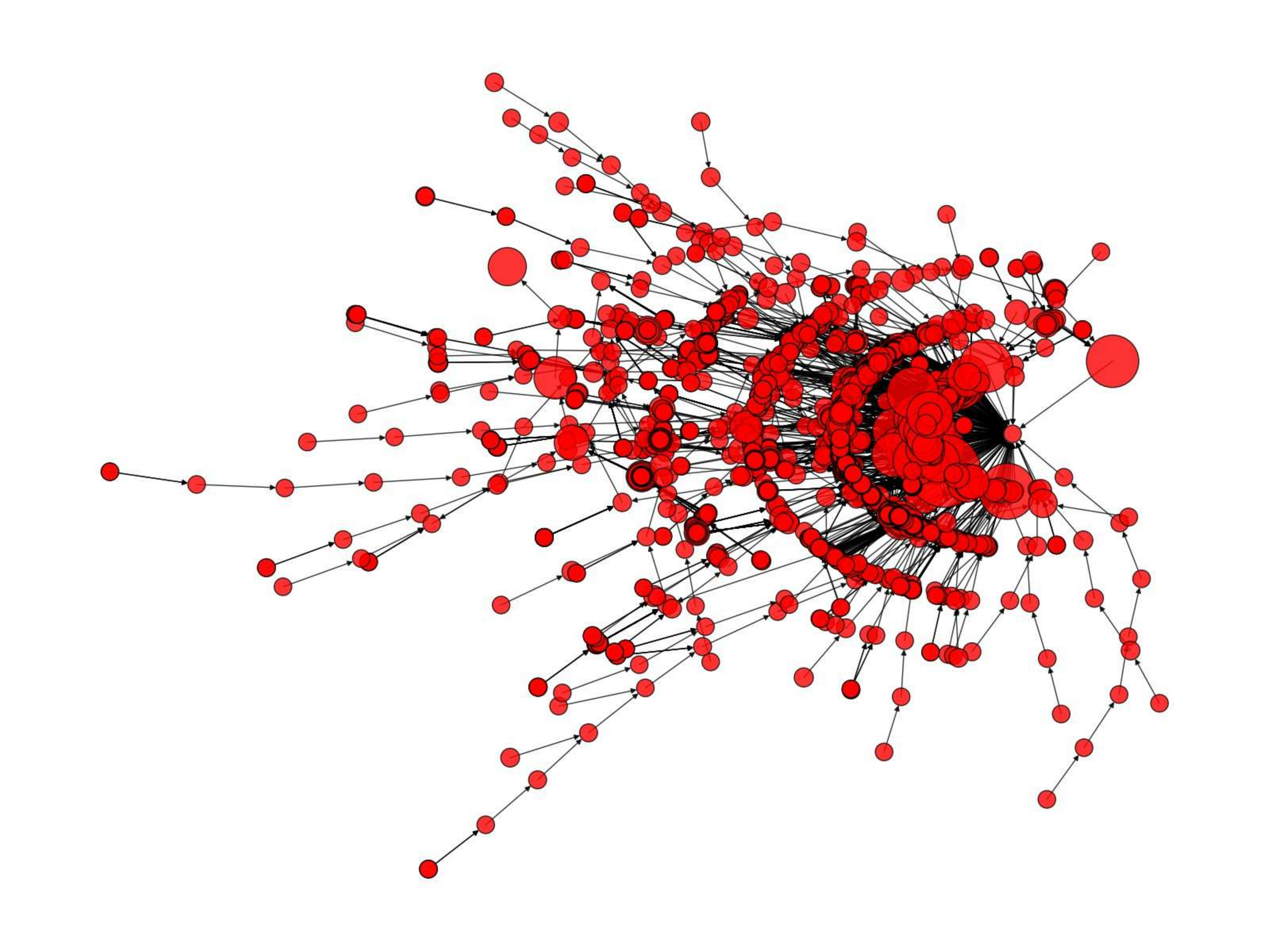}
        \label{fig:network-nonclickbait-size}
     }
     \caption{Endorsement Feature Path. The size of each node represents the endorsement attribute of the corresponding comment, i.e., the number of likes a comment received.}
     \label{fig:network-size}
 \end{figure*}

The Endorsement Path $RW_{e}$ captures the endorsement a comment receives in the comment network. It reflects the agreement from other users on a particular comment. Figure \ref{fig:network-size} demonstrates an example of the endorsement attribute in the comment network. We observe that only a few comments in clickbait videos receive a large amount of endorsements from other users. We find these comments are often the ones that point out the video is a clickbait, which are appreciated/endorsed by other users. The degree of endorsements in non-clickbait videos is much more diversified since users are more engaged in the discussion of the video content instead of endorsing or disputing it.    
 
We perform the random walk process $M$ times with a maximum length of $K$ for each attribute path, and denote each path with $RW_{s}(m)$ and $RW_{e}(m)$ for the $m^{th}$ walk of the \textit{sentiment} and \textit{endorsement} attribute respectively. If attribute paths $RW_s(m)$ and $RW_e(m)$ arrive at the source node $s$ before the length of the path reaches $K$, i.e., $v_k = s$ and $k < K$, $RW_s(m)$ and $RW_e(m)$ are padded with neutral sentiment and zero endorsement respectively. We store the $K$ attributes recorded in $M$ random walk paths into feature matrices ($H_s$ and $H_e$) of size $M \times K$, where $H_{s}(m, k) = RW_{s}(m, k)$ and $H_{e}(m, k) = RW_{e}(m, k)$.

\subsubsection{Network Representation Learning}

Given the high-dimensional network features of the comment network $\mathbf{G}$, namely $H_{s}$ and $H_{e}$ extracted by the Random Walk algorithm, we further encode the network features into a low-dimensional vector space and learn the latent vector representation for them. An autoencoder is a generative deep learning technique that captures hidden patterns of high dimensional data with artificial neural networks \cite{vincent2008extracting}. It consists of an \textit{encoder} $E(x): \mathbb{R}^n \rightarrow \mathbb{R}^m$ that reduces the feature space dimension from $n$ to $m$, where $m \ll n$. The reduction is done through one or more layers of neural networks, and a \textit{decoder} $D(z): \mathbb{R}^m \rightarrow \mathbb{R}^n$ that recovers the encoded vector to its original dimensionality. 

It has been shown that deep autoencoder is an effective way of nonlinear dimensionality reduction \cite{hinton2006reducing}. In our scheme, we compress the network feature vectors $H_{s}$ and $H_{e}$ by independently training two stacked autoencoders with respect to the sentiment and endorsement attributes. Formally, our stacked autoencoder structure consists of six neural network layers, and the $i^{th}$ layer is defined as: 
\begin{equation}
    \mathbf{Z}^i = \phi (\mathbf{W}^i \cdot \mathbf{X}^i + \mathbf{b}^i)
\end{equation}
where $\phi(\cdot)$ is a nonlinear activation function, $\mathbf{W}^i$ is the weighting factor and $\mathbf{b}^i$ is the bias term. $\mathbf{X}^i$ and $\mathbf{Z}^i$ are the input and output of each layer. We set the input of the first layer to be the network feature vectors $H_{s}$ and $H_{e}$ for the sentiment and endorsement autoencoder, respectively. We use Mean Square Error (MSE) as the loss function for the stacked autoencoder and the rectified linear unit (ReLU) as the activation function.

Finally, the latent features (i.e., $\mathbf{Z}_{s}$ and $\mathbf{Z}_{e}$) learned from the stacked autoencoders are aggregated by a concatenation function, $\mathbf{Z} = <\mathbf{Z}_{s}, \mathbf{Z}_{e}>$, which represents the network features extracted from the comment network.

\subsection{Linguistic Feature Extraction}\label{sec:linguistic}

\begin{figure*}[!htb]
\centering
     \subfigure[][Clickbait]{
         \centering
         \includegraphics[width=0.42\textwidth]{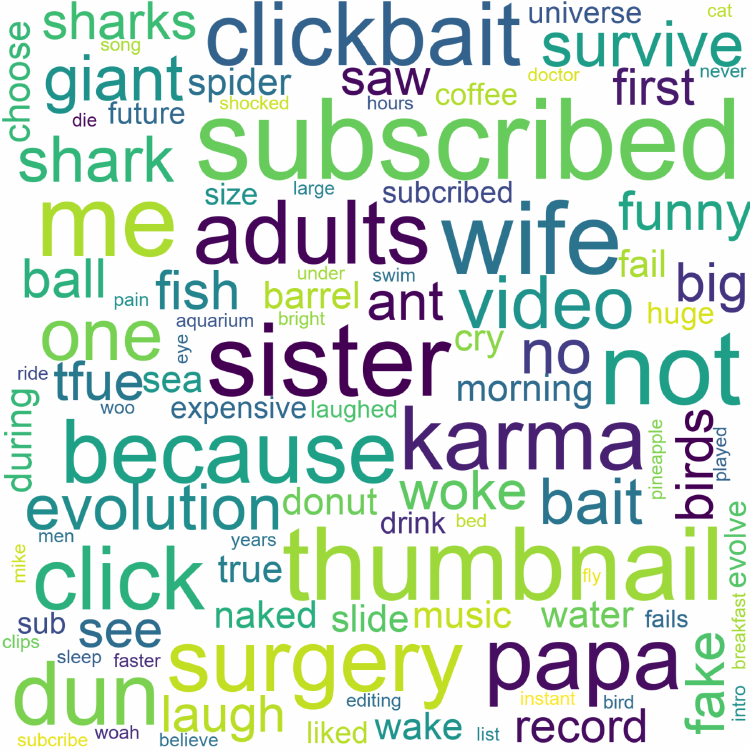}
         \label{fig:worldcloud-clickbait}
     }
     \subfigure[][Non-clickbait]{
         \centering
         \includegraphics[width=0.42\textwidth]{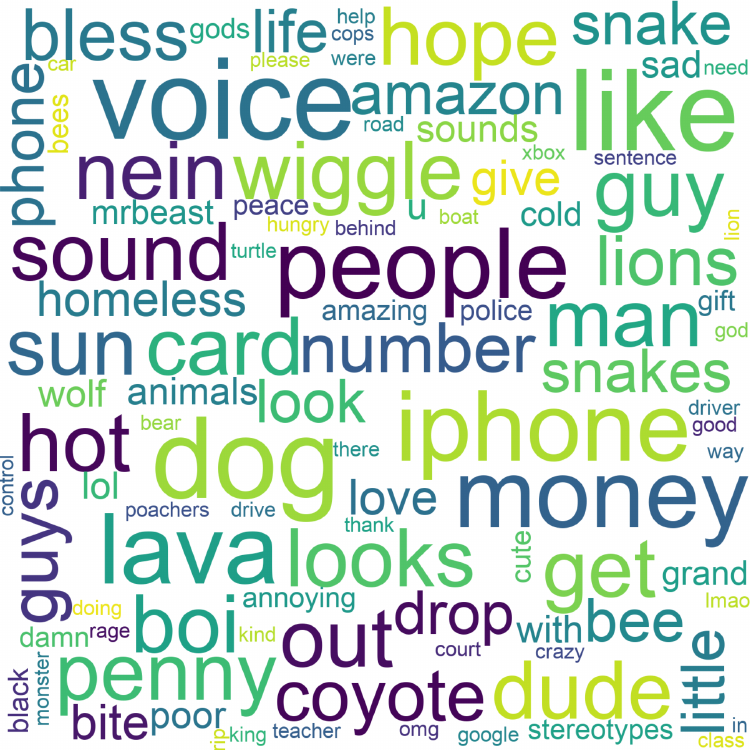}
        \label{fig:worldcloud-nonclickbait}
     }
     \caption{Word Clouds}
     \label{fig:wordcloud}
 \end{figure*}

In the second component, we extract linguistic features from the comment section by learning the document embedding of each comment. Figure \ref{fig:wordcloud} shows the word clouds of most frequent words from the comments of clickbait and non-clickbait videos, respectively. Intuitively, we observe that clickbait related words (e.g., ``clickbait", ``bait", ``fake'', ``thumbnail") appear more frequently in the comments of clickbait videos while content-related words (e.g., ``lava'', ``card'', ``voice'') appear more often in the discussion of non-clickbait videos. 

Therefore, we employ a widely adopted document embedding technique, namely Doc2vec \cite{Le2014doc2vec}, to extract linguistic features from comments (i.e., comment embedding). Doc2vec, derived from the famous Word2vec framework, is designed to learn fixed-length continuous distributed vector representations for word sequences of variable-length. 

A na\"ive approach is to simply embed the whole comment section of each video. However, we found such an approach performs poorly due to the extremely long or short comment length of some videos.  In particular, we found a majority of comment threads only have one comment, but some threads have hundreds of comments. To capture the information of both short and long comment threads, we design a hierarchical embedding process as follows: 1) we first embed each single comment using the Doc2vec model; 2) each long comment thread is traversed using the Random Walk process and the recorded comment path is further embedded by calculating the mean of comment embeddings in the path (referred to as \textit{path embedding}); 3) we train another independent 8-layer stacked autoencoder to reduce the dimension of the path embeddings. Using this hierarchical embedding method, we can control the dimensionality of the linguistic feature space while preserving the information captured by the Doc2vec framework.

\subsection{Metadata Feature Extraction} \label{sec:meta}

\begin{figure}[htb!]
    \centering
    \includegraphics[width=0.8\linewidth]{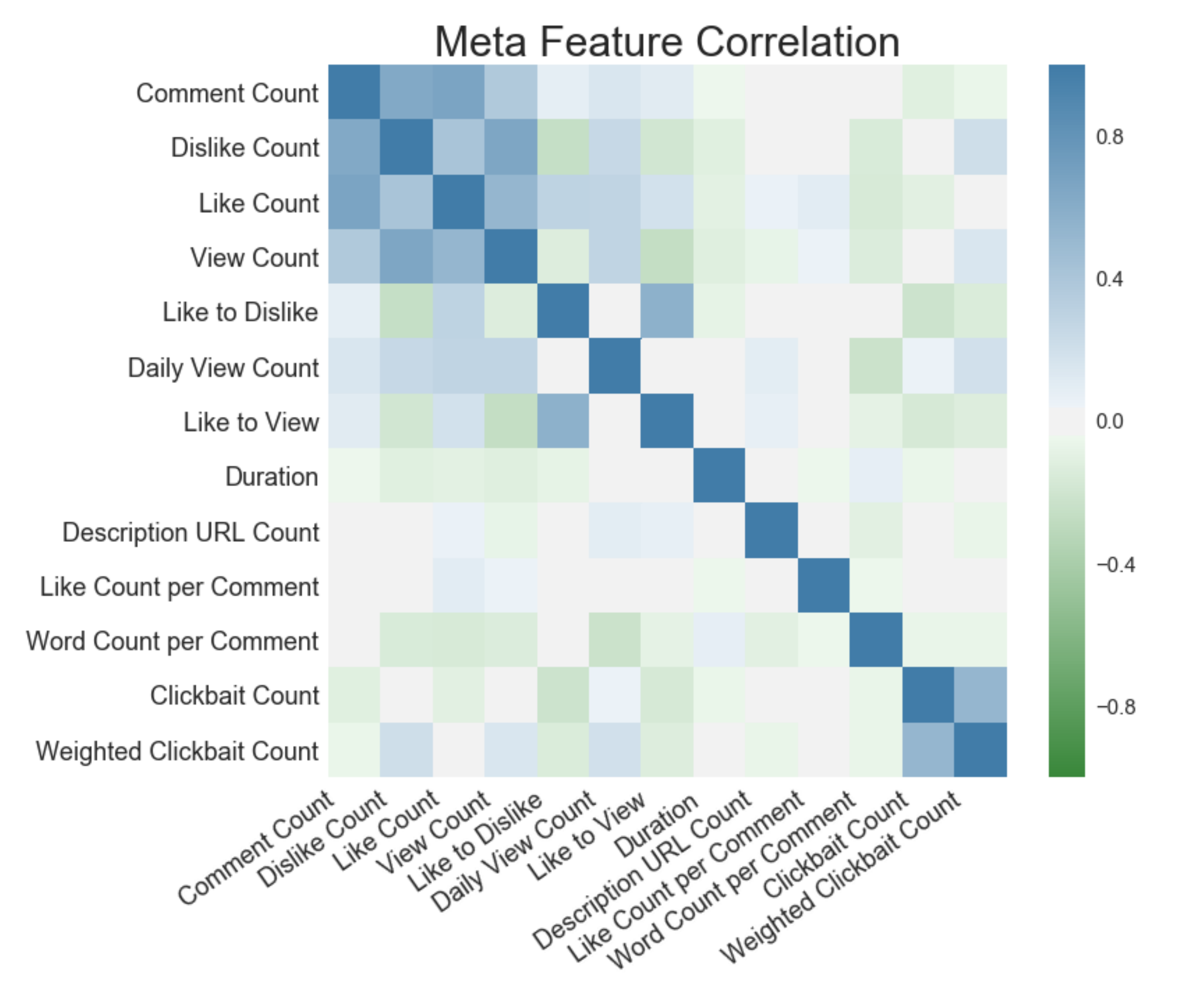}
    \caption{Metadata Feature Correlation}
    \label{fig:correlation}
\end{figure}

 We further extract some complementary metadata features that are relevant in identifying clickbait videos but cannot be captured by \lanyu{the} network and linguistic clues discussed above. These metadata features are mainly selected based on empirical observations. For example, we observe that some clickbait videos contain URLs to malicious websites in the video descriptors.  We extract a total of 13 metadata features from the collected real-world dataset. These metadata features are shown in Table \ref{tab:metafeatures}. These features are computed to describe statistical characteristics of both the video content (e.g., video length, the number of views) and the comments of the video (e.g., the word count of a comment, the number of likes of a comment).  The correlation plot of all the metadata features is shown in Figure \ref{fig:correlation}. We observe that the extracted metadata features are relatively independent.

\begin{table}[htb!]
    \small
    \caption{Metadata Features}
    \label{tab:metafeatures}
    \centering
    \begin{tabular}{l p{5.5cm}}
        \toprule
        \midrule
        \textbf{Feature} &  \textbf{Description}  \\
        \toprule
        \texttt{Comment Count} & Total \# of comments for each video \\
        \hline
        \texttt{Dislike Count} & Total \# of dislikes for each video \\
        \hline
        \texttt{Like Count} & Total \# of likes for each video  \\
        \hline
        \texttt{View Count} & Total \# of views for each video\\
        \hline
        \texttt{Like to Dislike} & The ratio of like count to dislike count \\
        \hline
        \texttt{Daily View Count} & Avg. \# of daily views for each video\\
        \hline
        \texttt{Like to View } & The ratio of like count to view count \\
        \hline
        \texttt{Duration} & Length of video in minutes \\
        \hline
        \texttt{Description URL Count} & Avg. \# of URLs in the description \\
        \hline
        \texttt{Like Count per Comment} & Avg. \# of likes for each comment \\
        \hline
        \texttt{Word Count per Comment} & Avg. \# of words in each comment \\
        \hline
        \multirow{2}{*}{\texttt{Clickbait Count}} & Avg. \# of words related to clickbait in each comment \\
        \hline
        \multirow{3}{*}{\texttt{Weighted Clickbait Count}} & Avg. \# of words related to clickbait in each comment weighted by comment's like count \\
        \midrule
        \toprule
    \end{tabular}
\end{table}

\subsection{Supervised Classification} \label{sec:classifier}
Finally, we combine all the network, linguistic and metadata features described above and perform the binary classification to detect online clickbait videos. In the supervised classification module, we adopt a few state-of-the-art classification algorithms and select the best-performed one as the classifier for our OVCP scheme. The supervised classifiers include probabilistic classifiers, support vector machine, boosting and ensemble methods, and neural networks. The detailed performance evaluation for the collection of classifiers and the OVCP scheme is presented in the following section.

\section{Evaluation} \label{sec:eval}
In this section, we first describe the dataset we collected from YouTube. We then evaluate the performance of the OVCP scheme in comparison with state-of-the-art baselines on the collected dataset. The results show that OVCP significantly outperforms both the compared baseline methods and human annotators in terms of accurately detecting online clickbait videos. 

\subsection{Dataset}
\label{subsec:dataset}
YouTube is visited by over 1.9 billion logged-in users each month and over a billion hours of video are watched daily\footnote{\url{https://www.youtube.com/intl/en-US/yt/about/press/}}. In view of the diversity and popularity of \lanyu{YouTube videos}, we take YouTube as our data source to collect video information.

For the training dataset, considering the imbalanced nature of clickbait videos on YouTube, we leverage YouTube's recommendation system to efficiently collect clickbait videos being recommended to real users and to minimize human bias in the data collection process. First, we randomly assigned a set of trending videos recommended by YouTube to three independent human annotators who collectively identified a set of ``seed'' clickbait videos by manually watching them (we set the size of the seed set to be 40 in our experiment due to the high labeling cost). We then created a dummy account on YouTube and let YouTube recommend relevant videos based on the watching history of the ``seed'' videos. Using such a method, we collected a training dataset of 500 videos which is a reasonable sample size  to study the online video detection problem on YouTube \cite{bajaj2016disinformation}. 
Note that, in the training set, we intentionally collect more clickbait videos so that the training set is relatively balanced. This allows us to train the model more effectively \cite{liu2008exploratory}. To ensure fairness, we use the same training data for all baselines. In order to evaluate the performance in a real-world scenario, we randomly collected 125 videos from the ``trending video” section \lanyu{on} YouTube’s homepage. This strategy allows the collected videos to have a similar clickbait video percentage as that on YouTube due to its nature of random sampling. In particular, following the common practice in supervised machine learning  \cite{domingos2012few} and the Pareto principle (i.e., 80/20 rule)\footnote{\url{https://en.wikipedia.org/wiki/Pareto\_principle}} of training/testing data split, we selected 500 videos as the training set and 125 videos as the testing set, and performed 5-fold cross-validation in our evaluation.

\begin{figure}[!htb]
    \centering
    \includegraphics[width=0.8\linewidth]{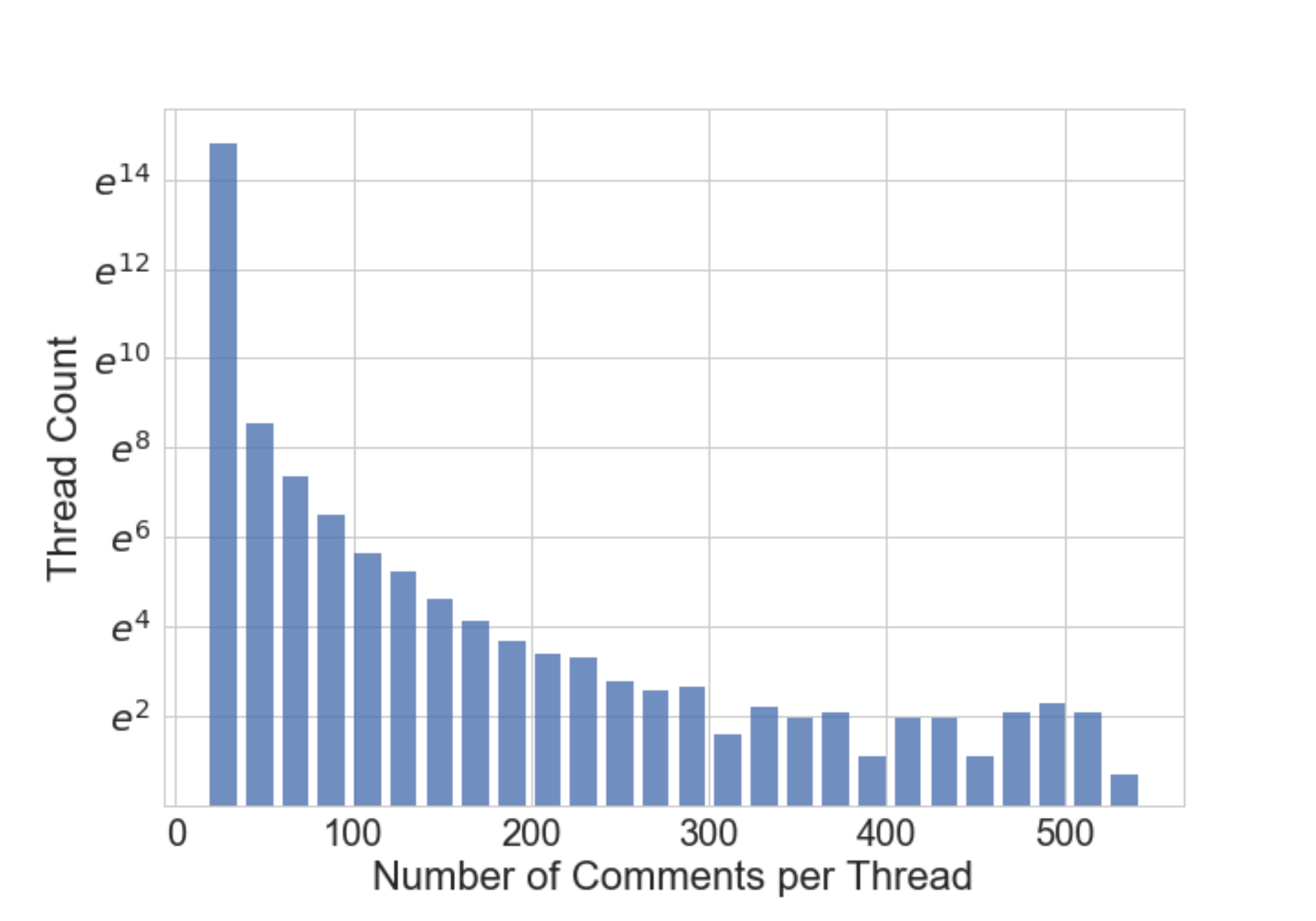}
    \caption{Distribution of Comments Count per Thread}
    \label{fig:comment_count}
\end{figure}

For each of these videos, we recorded its unique \texttt{videoID} and collected the ground-truth label (i.e., whether the video is a clickbait or not) based on the majority voting of human annotations. We also collected the title, description,  thumbnail,  comments  and commentThreads~\footnote{An attribute that records the structure of each comment thread in the comment section}, and the content of each video through the YouTube Data API\footnote{\url{https://developers.google.com/youtube/v3/}}. 
An overview of the collected dataset is summarized in Table \ref{table:datatrace}. We observe that the comments in the collected videos follow a ``long tail'' distribution. As shown in Figure \ref{fig:comment_count}, most comments in a thread have no replies (i.e., number of comments per thread is 1) and only a small portion of the threads have multiple comments.

\begin{table}[htb!]
    \caption{Data Trace Statistics}
    \label{table:datatrace}
    \small
    \centering
    \scalebox{0.88}{
    \begin{tabular}{l c c c c}
        \toprule
        \midrule
        \textbf{Dataset} &  \multicolumn{2}{c}{\textbf{Train}} & \multicolumn{2}{c}{\textbf{Test}}  \\
        \toprule
        \textbf{Video Label} &  \textbf{Clickbait} & \textbf{Non-clickbait} & \textbf{Clickbait} & \textbf{Non-clickbait}  \\
        \toprule
        Videos & 128 & 372 & 13 & 112\\
        \hline
        Comment Threads & 388,337 & 2,050,438 & 50,139 & 243,213 \\
        \hline
        Comments & 506,407 & 2,771,672 & 59,523 & 287,045\\
        \hline
        Comments per Thread & 1.30 & 1.35 & 1.19 & 1.18\\
        \hline
        Distinct Users & 419,392 & 2,072,927  & 51,998 & 246,481\\
        \hline
        Unique Words in Comments & 156,201 & 551,552 & 37,330 & 119,270 \\
        \midrule
        \toprule
    \end{tabular}
    }
\end{table}

\subsection{Baselines}
We integrate the following state-of-the-art supervised classifiers \cite{alpaydin2014introduction} with our scheme as discussed in Section~\ref{sec:classifier}: \textit{Logistic Regression (LR)}, \textit{Linear Support Vector Machine (SVM)}, \textit{AdaBoost}, \textit{Random Forest (RF)}, and \textit{Multi-layer Perceptron (MLP)}. We select the best-performed one to be the one used in our OVCP scheme.  We compare the performance of our scheme with the following representative baselines in clickbait detection \footnote{The online video clickbait detection problem is a largely unsolved research problem. Therefore, we only have a limited number of baselines from the literature to compare against.}. 

\begin{itemize}
\item \textbf{Image-based Clickbait Detection (VGG-16):} A convolutional neural network approach (pre-trained on ImageNet) that detects the clickbait video using only the image content of the thumbnail \cite{zhang2015accelerating}.

\item \textbf{Stop Clickbait (ASONAM16):}  A linguistic-based clickbait classification approach that detects the clickbait video based on the title content of the video \cite{chakraborty2016stop}.

\item \textbf{Clickbait Detection using Deep Learning (NGCT16):} A deep neural network approach that identifies clickbait videos using a compiled clickbait corpus from social media posts\cite{agrawal2016clickbait}.

\item \textbf{Clickbait Video Detection (SPW18):} A deep learning based approach that detects clickbait video on YouTube using features from the headline, thumbnail, comments and video statistics \cite{zannettou2018good}.
\end{itemize}

\subsection{Detection Accuracy}
In the first set of experiments, we evaluate the performance of our scheme when it is coupled with a set of classifiers and identify the best-performed one as the OVCP scheme. The evaluation metrics involved in measuring the effectiveness of clickbait video detection include \textit{Accuracy}, \textit{Precision}, \textit{Recall}, and \textit{F1-Score}, which are commonly used in \lanyu{evaluating} binary classification tasks. In addition,  we include two imbalanced metrics, \textit{Cohen's Kappa (Kappa)} and \textit{Matthews Correlation Coefficient (MCC)}, because our dataset is not perfectly balanced (see Table~\ref{table:datatrace}). For all classification methods and baselines, we perform 5-fold cross-validation in the parameter tuning process.

In the  OVCP scheme, we set the parameters in each component primarily based on our empirical observation as follows: i) \textit{random walk}: we set $M=100$ and $K=5$;  ii) \textit{stacked autoencoder (network)}: we set the size of encoding layers to be 256, 64 and 16, and the size of decoding layers to be 64, 256 and 500; iii) \textit{Doc2vec}: the length of the comment embedding is set to be 256; and iv) \textit{stacked autoencoder (linguistic)}: we set the size of encoding layers to be 128, 64, 32 and 16, and the size of decoding layers to be 32, 64, 128 and 256. 

The performance of all compared schemes is summarized in Table \ref{tab:classification}. We observe that AdaBoost performs the best among all classifiers and thus is selected to be the default classifier in the OVCP scheme. We also observe that our scheme outperforms all baseline approaches \lanyu{under} all evaluation metrics. In particular, comparing to the \textit{VGG-16}, \textit{ASONAM16}, \textit{NGCT16} and \textit{SPW18} baselines, our scheme achieves a performance gain of 0.3262, 0.2219, 0.1943 and 0.1229 on the F1-score respectively. Although the \textit{SPW18} approach considers various content-based features (e.g., headline, thumbnail, video statistics) in its model, it is found to be less effective than our scheme in detecting clickbait videos.  The reason is that the \textit{SPW18} method has a high-dimensional input vector size of 855 that requires a vast amount of training samples to achieve a reasonable performance. The \textit{SPW18} approach also relies on visual features (i.e., thumbnail) which can be misleading in detecting the clickbait videos. 

\begin{table}[htb!]
  \small
  \centering
  \caption{Clickbait Classification Performance for All Methods}
  \scalebox{0.88}{
  
    \begin{tabular}{l c c c c c c}
    \toprule
    \midrule
    \textbf{Algorithms}  &\textbf{Accuracy}&\textbf{Precision}&\textbf{Recall}&\textbf{F1-Score}&\textbf{Kappa} & \textbf{MCC}\\
    \toprule
    AdaBoost (\textbf{OVCP}) & \textbf{0.8960 } & \textbf{0.5000} & \textbf{0.4615} & \textbf{0.4800} & \textbf{0.4223} &\textbf{ 0.4227} \\
    \cmidrule(l){1-7}
    LR & 0.8800 & 0.4286  & 0.4615 & 0.4444 & 0.3773  & 0.3776\\ 
    \cmidrule(l){1-7}
    SVM & 0.8880  & 0.4545 & 0.3846 & 0.4167 & 0.3552 & 0.3567 \\
    \cmidrule(l){1-7}
    RF  & 0.8640 & 0.3889 & 0.5385 & 0.4516 & 0.3763 & 0.3828 \\
    \cmidrule(l){1-7}
    MLP        & 0.8480 & 0.2857 & 0.3077 & 0.2963  & 0.2112 & 0.2114\\
    \cmidrule(l){1-7}
    \textbf{VGG-16} & 0.8240 & 0.1538 & 0.1538 & 0.1538 & 0.0556 & 0.0556 \\
    \cmidrule(l){1-7}
    \textbf{ASONAM16} & 0.8160 & 0.2222 & 0.3077 & 0.2581 & 0.1561 & 0.1588\\
    \cmidrule(l){1-7}
    \textbf{NGCT16} & 0.8400 & 0.2667 & 0.3077 & 0.2857 & 0.1961  & 0.1968\\
    \cmidrule(l){1-7}
    \textbf{SPW18} & 0.8560 & 0.3333 & 0.3846 & 0.3571 & 0.2765  & 0.2774\\
    \midrule
    \toprule
    \end{tabular}
    }
\label{tab:classification}
\end{table}

\begin{figure}[!htb]
    \centering
    \includegraphics[width=0.8\linewidth]{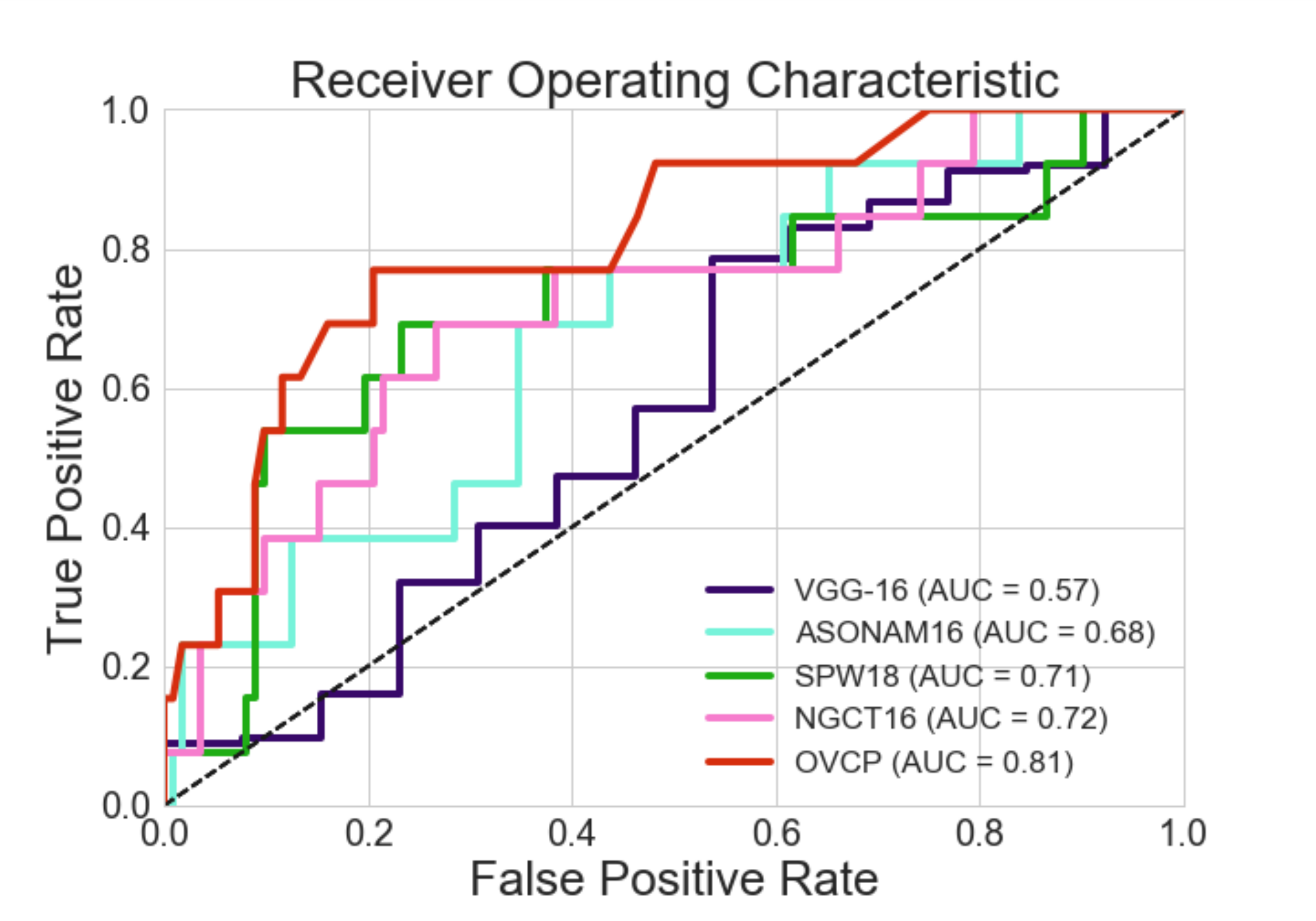}
    \caption{ROC Curve of All Schemes}
    \label{fig:roc}
\end{figure}

Additionally, we plot the Receiver Operating Characteristic (ROC) curve of all schemes to evaluate the robustness of their performance with respect to the classification threshold. As shown in Figure \ref{fig:roc}, our scheme consistently outperforms all baselines and achieves an increase of 0.09 on the AUC score compared to the baseline of best performance (i.e., NGCT16).   

\subsection{Feature Analysis}
In the second experiment, we study the importance of features in each category (i.e., network, linguistic, metadata) and their combinations. The results are shown in Table \ref{tab:classification_feature_set}. We observe that the combination of all features achieve the best results on all evaluation metrics which confirms their necessity in our model. We also observe that linguistic features achieve the best performance among all single feature categories, and the combination of linguistic and metadata features achieve the best performance among the combination of any two feature categories.

\begin{table}[htb!]

  \small 
  \centering
  
  \caption{Classification Performance for All Feature Sets}
  \vspace{-0.1in}
  \scalebox{0.88}{
    
    \begin{tabular}{l c c c c }
    \toprule
    \midrule
    \textbf{Feature Set}  &\textbf{Accuracy}&\textbf{Precision}&\textbf{Recall}&\textbf{F1-Score} \\
    \toprule
    \textbf{All Features}& \textbf{0.8960 } & \textbf{0.5000} & \textbf{0.4615} & \textbf{0.4800} \\
    \cmidrule(l){1-5}
    \textbf{Metadata}  &0.7680 &0.1364 &0.2308 &0.1714  \\ 
    \cmidrule(l){1-5}
    \textbf{Network} &0.8240 &0.1538 &0.1538 &0.1538 \\
    \cmidrule(l){1-5}
    \textbf{Linguistic} &0.8480 &0.2000 &0.1538 &0.1739 \\
    \cmidrule(l){1-5}
    \textbf{Network \& Metadata} &0.8720 &0.2857 &0.1538 &0.2000 \\
     \cmidrule(l){1-5}
    \textbf{Linguistic \& Metadata } &0.8400 &0.2941 &0.3846 &0.3333\\
    \cmidrule(l){1-5}
    \textbf{Network \&Linguistic}  &0.8560 &0.2727 &0.2308 &0.2500 \\
    \midrule
    \toprule
    \end{tabular}
    }
\label{tab:classification_feature_set}
\end{table}

\subsection{Influence of Training Size}

\begin{figure}[!htb]
    \centering
    \includegraphics[width=0.8\linewidth]{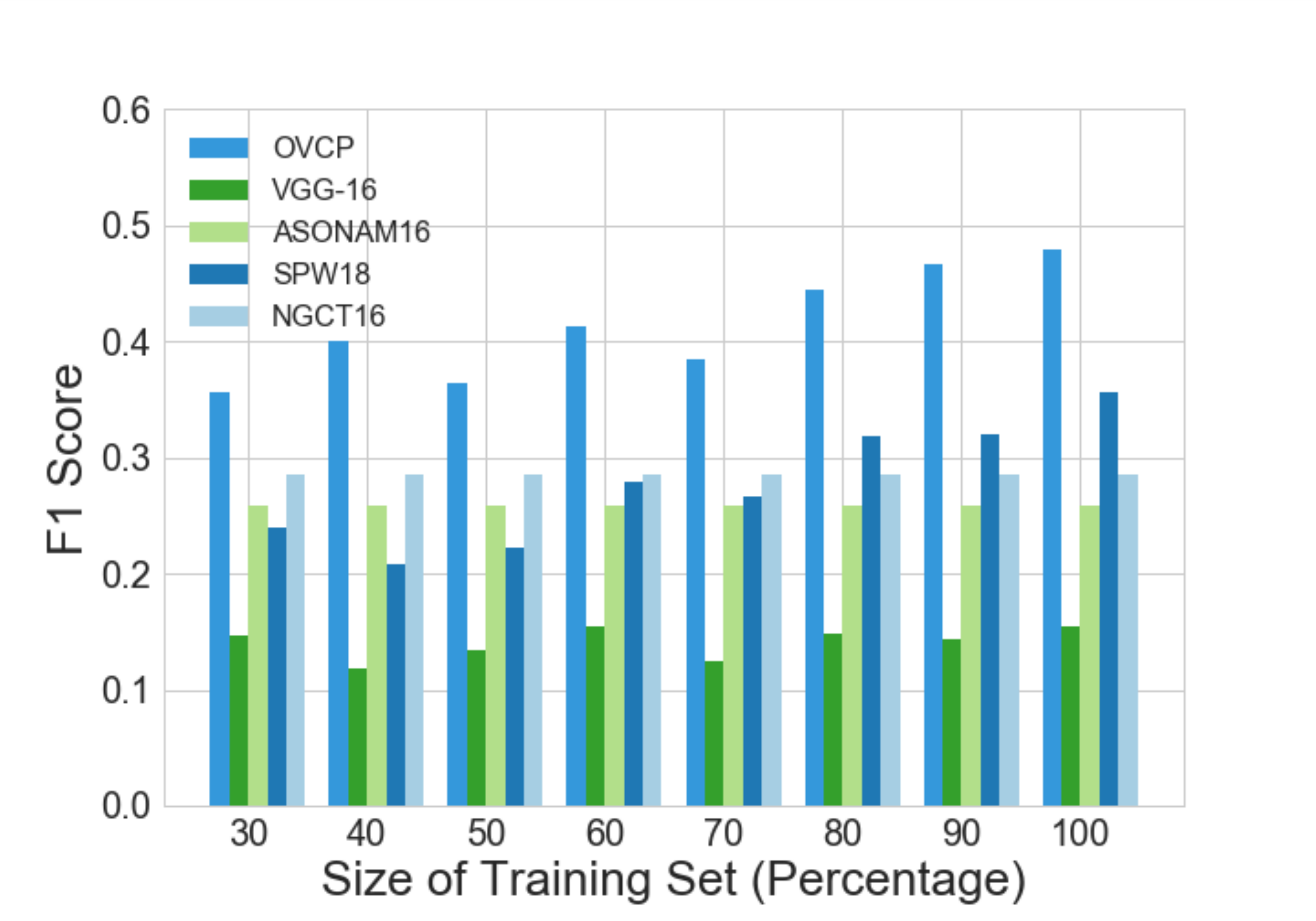}
    \caption{Size of Training Set vs. Performance (F1 Score)}
    \label{fig:trainsize}
\end{figure}

In the third experiment, we further study the robustness of our scheme against the size of the training set. In particular, we evaluate the performance of the OVCP scheme and all other baselines by increasing the size of the training set from 30\% to 100\% of the entire \lanyu{training} set. The F1-score results are reported in Figure \ref{fig:trainsize}.  We observe that the performance of our scheme generally improves as the size of the training set increases and continues to outperform other baselines.

\subsection{Detection Time}

In the fourth experiment, we measure the detection performance of the OVCP scheme against the time frame after the video is posted. Specifically, we limit the comment data in our scheme to be within a time frame from the corresponding video’s publishing time, and increase the time frame from 10 minutes to 24 hours. The classification results are summarized in Table \ref{tab:classification_time}. We observe that the OVCP scheme using \lanyu{comments within} the first 10 minutes after the video’s publication has already outperformed all content-based baselines with a non-trivial performance gain. Moreover, the OVCP scheme is observed to reach its optimal performance within 24 hours.

\begin{table}[htb!]
  \small 
  \centering
  
  \caption{Classification Performance v.s. Detecting Time Frame}
  \vspace{-0.1in}
  \scalebox{0.88}{
    
    \begin{tabular}{l c c c c }
    \toprule
    \midrule
    \textbf{Time Frame}  &\textbf{Accuracy}&\textbf{Precision}&\textbf{Recall}&\textbf{F1-Score} \\
    \toprule
    \textbf{OVCP (Within 10 Minutes)}  & 0.8640 & 0.3571 & 0.3846 & 0.3704 \\ 
    \cmidrule(lr){1-5}
    \textbf{OVCP (Within 30 Minutes)}  & 0.8560 & 0.3684 & 0.5385 & 0.4375  \\ 
    \cmidrule(lr){1-5}
    \textbf{OVCP (Within 1 Hour)}  & 0.8640 & 0.3889 & 0.5385 & 0.4516 \\ 
    \cmidrule(lr){1-5}
    \textbf{OVCP (Within 6 Hours)} & 0.8720 & 0.4118 & 0.5385 & 0.4667 \\
    \cmidrule(lr){1-5}
    \textbf{OVCP (Within 12 Hours)} & 0.8880 & 0.4615 & 0.4615 & 0.4615 \\
    \cmidrule(lr){1-5}
    \textbf{OVCP (Within 24 Hours)} & 0.8960 & 0.5000 & 0.4615 & 0.4800 \\
    \cmidrule(lr){1-5}
    \textbf{OVCP (All Comments)}  &\textbf{0.8960 } & \textbf{0.5000} & \textbf{0.4615} & \textbf{0.4800} \\
    \cmidrule(lr){1-5}
    \textbf{VGG-16} & 0.8240 & 0.1538 & 0.1538 & 0.1538  \\
    \cmidrule(lr){1-5}
    \textbf{ASONAM16} & 0.8160 & 0.2222 & 0.3077 & 0.2581 \\
    \cmidrule(lr){1-5}
    \textbf{NGCT16} & 0.8400 & 0.2667 & 0.3077 & 0.2857 \\
    \cmidrule(lr){1-5}
    \textbf{SPW18} & 0.8560 & 0.3333 & 0.3846 &0.3571\\
    \midrule
    \toprule
    \end{tabular}
    }
\label{tab:classification_time}
\end{table}

\subsection{Detection Efficiency}

\begin{figure}[!htb]
    \centering
    \includegraphics[width=0.78\linewidth]{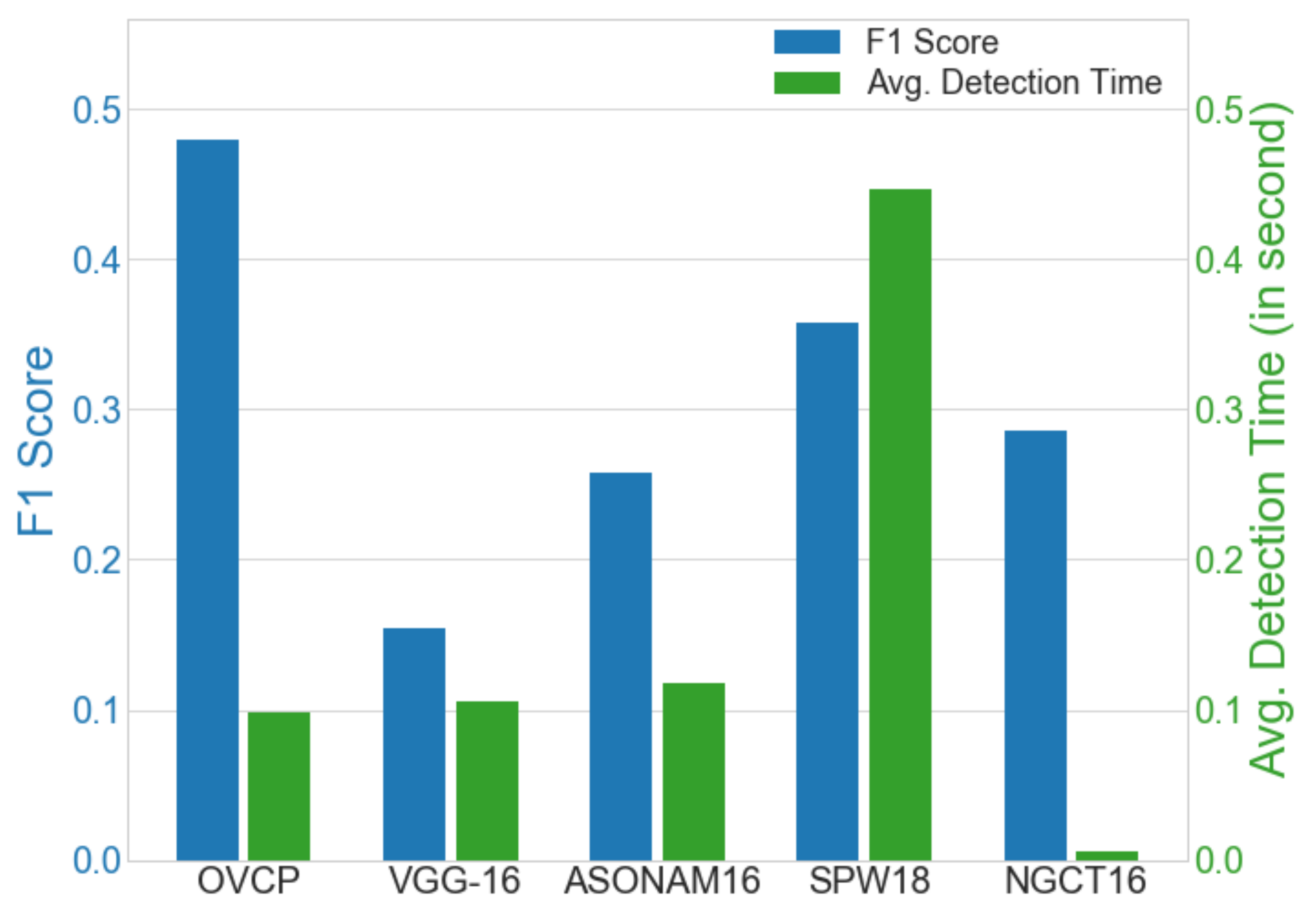}
    \caption{Performance (F1 Score) v.s. Average Detection Time Cost (per Video)}
    \label{fig:detectiontime}
\end{figure}

In the fifth experiment, we empirically measure the efficiency (i.e., detection time per video) of all schemes. The results are shown in Figure~\ref{fig:detectiontime}. We observe that the OVCP scheme achieves the best detection performance (F1 score) with a reasonable detection time compared to other baselines. We also observe that the SPW18 approach is the slowest among all compared schemes. This is because SPW18 requires features to be extracted from multiple data modalities, including image (i.e., thumbnail), texts (i.e., video title and description) and metadata. Such a large feature space significantly reduces the detection efficiency. The NGCT16 has the least detection time since it simply relies on pre-trained word embeddings to predict clickbait videos via a deep learning framework, which in turn limits its detection accuracy.

\begin{figure}[!htb]
    \centering
    \includegraphics[width=0.8\linewidth]{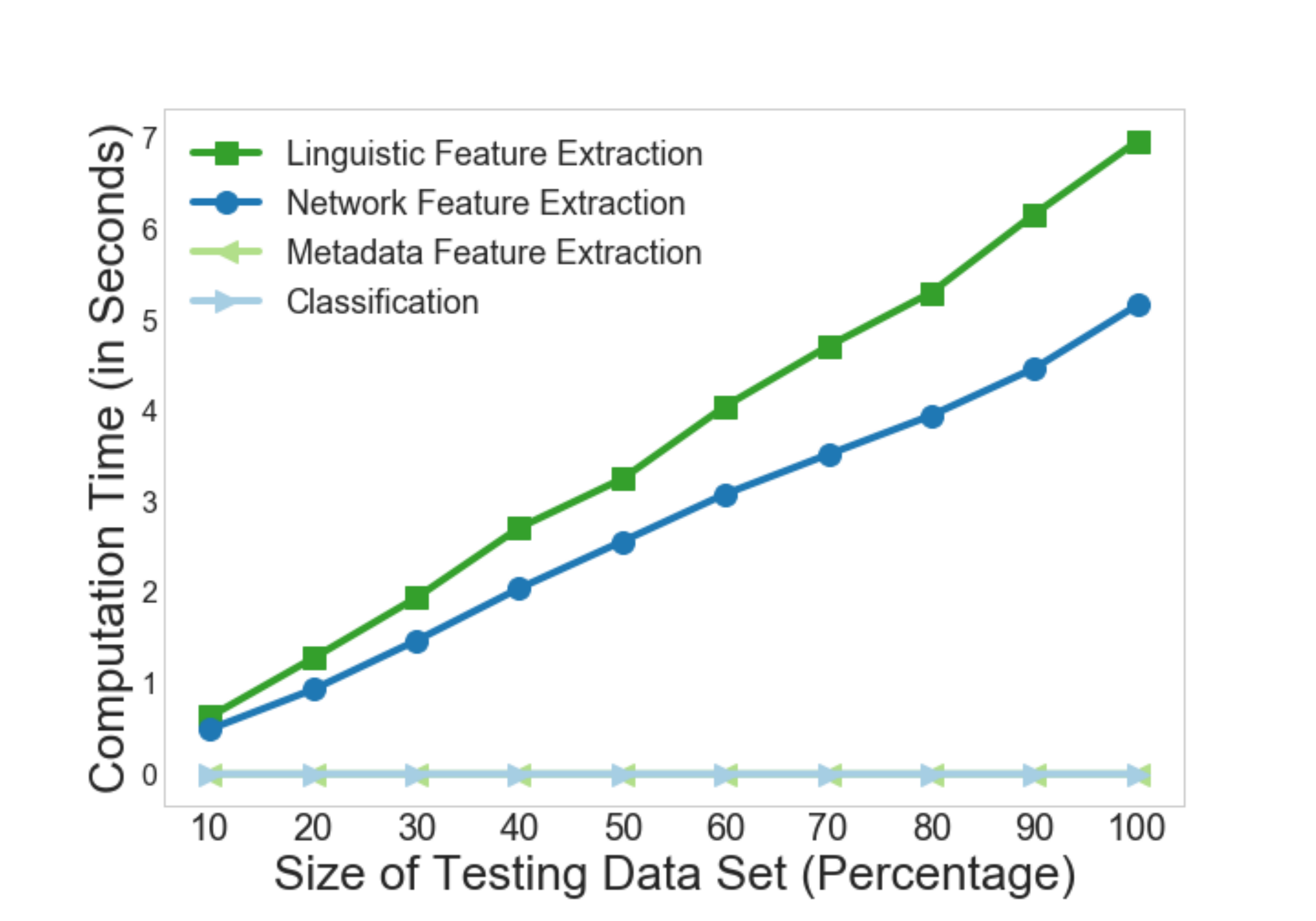}
    \caption{Computation Time for All Modules of OVCP}
    \label{fig:computingtime}
\end{figure}

Finally, we study the computational complexity/scalability of our OVCP scheme. In particular, we perform an empirical study of different modules of the OVCP scheme for a detailed analysis. The results are shown in Figure~\ref{fig:computingtime}. We observe that the linguistic and network feature extraction modules are the two dominant ones that consume most of the execution time of the OVCP scheme. The reason is that these dominant modules include neural network solutions (i.e., doc2vec and autoencoder) for latent feature extraction, which often involve a non-trivial amount of matrix operations in both learning and inference phases of the process \cite{hertz2018introduction}. However, we observe the computation time of those two dominant modules increase ~\emph{linearly} as the size of the input data increases, which demonstrates the good scalability of our scheme with larger datasets.

We also note that, in our current implementation, the videos are detected in a sequential manner, i.e., one video is detected after another. This leads to the linear growth of the execution time as the number of videos becomes large. We observe that our OVCP scheme can be easily extended to further improve its speed by leveraging the parallel GPU programming frameworks such as CUDA\footnote{\url{https://developer.nvidia.com/cuda-zone}} to execute many video streams in parallel on thousands of GPU cores. Moreover, elastic distributed computing systems such as AWS Kubernetes\footnote{\url{https://kubernetes.io/}} can also allow new OVCP instances to spin up on virtual machines when the system is overloaded (e.g., when many new videos are uploaded). Considering this line of effort is beyond the scope of this paper, we plan to implement it in our future work.

\subsection{Human Performance Comparison}

In the last experiment, we compare the performance of OVCP with human annotators\footnote{The difference between the human annotators in this experiment and the human annotators we used in collecting the ground truth labels is that we did not allow the human annotators in this experiment to watch the actual videos before they generate the annotations.}. In particular, we invited three independent annotators (different from the ones who were invited to generate the ground truth labels) to identify clickbait videos from the same testing dataset used for the OVCP scheme. First, we asked these annotators to annotate videos by only showing them the title, thumbnail and view count of each video which \lanyu{are} exactly the same information a YouTube user receives \textit{before} clicking a video. Next, we asked these annotators to annotate the videos by also giving them access to the comment section of the videos (i.e., metadata and comment features).
The performance of each individual annotator and the aggregated results using the majority voting (i.e., ``overall'' and ``overall+comment'') are shown in Table \ref{tab:human}. We also report false positive rate (FPR), and false negative rate (FNR) in addition to accuracy and F1 score metrics to study what kinds of mistakes the human annators made. We observe that the OVCP scheme consistently outperforms human annotators in detecting clickbait videos, demonstrating the necessity of developing such a tool. We also observe the human annotators often perform worse when they are not allowed to access the comments of the video, especially in terms of false negative rate (\lanyu{i.e.,} missing many clickbait videos). 

\begin{table}[htb!]
  \small  
  \centering
  \caption{Comparison between OVCP and Human Performance}
  \vspace{-0.1in}
  \scalebox{1}{

    \begin{tabular}{l c c c c }
    \toprule
    \midrule
     &\textbf{Accuracy}&\textbf{F1-Score}&\textbf{FPR}&\textbf{FNR}\\
    \toprule
    \textbf{OVCP} & \textbf{0.8960} & \textbf{0.4800} & \textbf{0.0536} & \textbf{0.5385}  \\
     \cmidrule(l){1-5}
     Annotator 1 (without comments) & 0.7040 & 0.0976 & 0.2321 & 0.8462 \\ 
     \cmidrule(l){1-5}
     Annotator 1 (with comments)    & 0.7440 & 0.1579 & 0.1964 & 0.7692 \\
     \cmidrule(l){1-5}
     Annotator 2 (without comments) & 0.6880 & 0.1333 & 0.2589 & 0.7692 \\ 
     \cmidrule(l){1-5}
     Annotator 2 (with comments)  & 0.7360 & 0.1951 & 0.2143 & 0.6923   \\
     \cmidrule(l){1-5}
     Annotator 3 (without comments) & 0.7520 & 0.1143 & 0.1786 & 0.8462  \\ 
     \cmidrule(l){1-5}
     Annotator 3 (with comments)  & 0.7920 & 0.1875 & 0.1429 & 0.7692  \\
     \cmidrule(l){1-5}
     Overall (without comments) & 0.7280  & 0.1053  & 0.2054  & 0.8462  \\ 
     \cmidrule(l){1-5}
     Overall (with comments) & 0.7680  & 0.1714  & 0.1696 & 0.7692  \\
    \midrule
    \toprule
    \end{tabular}
    }
\label{tab:human}
\end{table}

\section{Conclusion and Future Work}

In this paper, we develop a content-free scheme (OVCP) to detect clickbait videos on online video sharing platforms. Our scheme leverages the comment and interaction between users who watched the video, and learns latent features from their unstructured and complex comments. We evaluate our scheme using the real-world data collected from YouTube. The results demonstrate that our scheme is more accurate than both state-of-the-art clickbait detection tools and human annotators in identifying online clickbait videos. 
 
Finally, we also outline a few future research directions that can be built on the work from this paper. First, clickbait videos on YouTube are often customized by experienced content creators who know how to game with the YouTube recommendation system. To make the OVCP scheme more robust and generalizable across different video platforms, we plan to extend OVCP's compatibility with other video sharing platforms using different commentary structures. Examples of such platforms include both video-centric platforms (e.g., Vimeo) and social-based platforms (e.g., Twitter).  
Second, there exist many marketing vendors from which content creators can buy comments and high-retention views. In future work, we will further study how to identify and filter such fake or machine-generated comments. A possible solution is to leverage the social media bot detection techniques \cite{ferrara2016rise} and train a classifier that can discriminate comments generated by humans and bots. Meanwhile, we can also adopt the truth discovery \cite{zhang2018towards} and fact-checking \cite{wang2019age} approaches to verify the truthfulness of the user comments. The authors believe the above extensions will further improve the effectiveness and robustness of the OVCP scheme in detecting online clickbait videos.




\bibliographystyle{elsarticle-num}
\bibliography{refs}

\end{document}